\documentclass{article}

\usepackage{natbib}
\usepackage[dvips]{graphicx}
\usepackage{pstricks,pst-node,multido}
\usepackage{amsmath,amssymb}
\usepackage{url,subfigure}
\usepackage{ifthen}

\usepackage{ntheorem}
\newcounter{thm}
\newtheorem{lem}{Lemma}

\newtheorem{thm}[lem]{Theorem}
\newtheorem{cor}[lem]{Corollary}
\theoremstyle{plain}
\theoremseparator{~--}
\theorembodyfont{\rmfamily}
\theoremsymbol{\rule[1mm]{3.5mm}{1pt}}
\newtheorem{exm}{Example}
\newtheorem{rmk}[lem]{Remark}
\theoremstyle{plain}
\theoremsymbol{$\square$}
\newtheorem{prf}{Proof}

\newcommand{\Nat}{\mathbb{N}}
\renewcommand{\mod}{\textrm{ \small mod }}
\newcommand{\Aut}{\textrm{Aut}(\Gamma)}
\newcommand{\Tran}{\textrm{Trans}(\Gamma)}
\newcommand{\textcolor}[2]{#2}
\newcommand{\bigstrut}{\vphantom{\Big\{}}

\newrgbcolor{sky}{0.8 0.8 1}
\newrgbcolor{notwhite}{0.9 0.9 0.9}
\newrgbcolor{notgray}{0.63 0.63 0.63}
\newrgbcolor{lightvio}{0.5 0.5 0.5}
\newrgbcolor{lightblue}{0.4 0.4 1}
\newrgbcolor{turquoise}{0.5 0.5 0.5}
\newrgbcolor{gold}{0.7 0.7 0.7}
\newrgbcolor{lightgold}{1 0.9 0.3}
\newrgbcolor{violet}{0 0 0}
\newrgbcolor{green}{0.5 0.5 0.5}
\newrgbcolor{lightgreen}{0.75 0.75 0.75}
\newrgbcolor{lightred}{1 0.7 0.7}
\newrgbcolor{orange}{0.7 0.7 0.7}
\newrgbcolor{lightora}{0.95 0.95 0.95}

\begin{document}

\title{The vertex-transitive \\ TLF-planar graphs}

\author{David {Renault} \\
 \url{renault@labri.fr} \\[2mm] 
 LaBRI -- Universit\'e Bordeaux I \\
 351, cours de la Lib\'eration \\ 33400 Talence, \textsc{France}}

\maketitle

\begin{abstract}
  We consider the class of the topologically locally finite (in short
  TLF) planar vertex-transitive graphs, a class containing in
  particular all the one-ended planar Cayley graphs and the normal
  transitive tilings. We characterize these graphs with a finite local
  representation and a special kind of finite state automaton named
  \emph{labeling scheme}. As a result, we are able to enumerate and
  describe all TLF-planar vertex-transitive graphs of any given
  degree.  Also, we are able decide to whether any TLF-planar
  transitive graph is Cayley or not.\\[1mm]
  
\end{abstract}

\section*{Introduction}

Vertex-transitive graphs -- or transitive graphs in short -- are
graphs whose group of automorphisms acts transitively on their sets of
vertices. These graphs possess a regular structure, being structurally
the same from any vertex. When such a graph is planar, this regular
structure confers symmetry properties to the embedding of the graph~:
the action of automorphisms on the graph can locally be represented as
the action of an isometry of the geometry it is embedded in.

The class $\mathcal{T}$ of the topologically locally finite (in short
TLF) planar transitive graphs is a subclass of the class of the
transitive planar graphs. These graphs possess a planar embedding such
that the set of vertices in this embedding is a locally finite subset
of the plane. For example, $\mathcal{T}$ contains the planar Cayley
graphs of the discrete groups of isometries of the plane, be it
hyperbolic or Euclidean. We find in $\mathcal{T}$ both tree-like
graphs of finite treewidth and one-ended graphs such as the Euclidean
grid.

The TLF-planar graphs are related to several fields: first, such
graphs represent adapted models of computation for parallel algorithms
such as cellular automata \cite{Mazoyer,Garzon}, and provide examples
of structures for interconnection networks \cite{Heydemann}. The class
$\mathcal{T}$ is also connected to the vertex-transitive tilings of
the plane whose set of vertices is topologically locally finite
\cite{Tilings}. Second, while many problems in combinatorial group
theory are undecidable, the planar vertex-transitive graphs possess
structural as well as geometrical properties that allow for a more
specific approach. For example, the unique embedding property in the
case of the 3-connected planar graphs \cite{Whitney} constrains the
structure of the automorphisms of the graph.  Finally, putting aside
the finite graphs in $\mathcal{T}$, the infinite graphs in
$\mathcal{T}$ may factor into finite transitive graphs embedded into
compact manifolds, as it is the case for Euclidean
graphs~\cite{Bieberbach}.

The characterization of the finite planar transitive graphs is a
result of Fleischner and Imrich \cite{Fleischner}. These graphs turn
out to be the complexes associated to the uniform convex polyhedra.
The finite non-planar case happens to be much more difficult, and has
been thoroughly studied up to 26 vertices by McKay {\it et al.}
\cite{McKayRoyle,RoyleURL}, but few general results are known.  Cayley
graphs are natural examples of vertex-transitive graphs, and McKay
also studied the problem of the determination of those graphs that
were transitive but not Cayley graphs
\cite{McKayPraegerI,McKayPraegerII}. The problem of enumerating the
normal Cayley graphs, has been solved by Chaboud \cite{Chaboud} and
then extended to the TLF-planar Cayley graphs \cite{DavidCayley}.

In this paper, we give an exhaustive description of the class
$\mathcal{T}$ of the TLF-planar vertex-transitive graphs. Our
description of the class includes both finite and infinite graphs.
This extends the Cayley case \cite{DavidCayley} in several ways.
First, the Cayley case is mainly dedicated to the description of
groups which happen to have a planar Cayley graph. This article
focuses on the graphs themselves, by describing all their possible
groups of automorphisms, and as a consequence all their possible
embeddings. Second, we can highlight properties of the transitive
graphs that do not hold when we restrict ourselves to Cayley graphs.
For example, TLF-planar Cayley graphs can always be represented by the
Cayley graph of a discrete group of isometries of the plane. There
exist transitive graphs for which this is impossible, independently of
the embedding. Finally, $\mathcal{T}$ contains a strictly larger class
of graphs than the Cayley case. Such a simple graph as the complex
associated to the dodecahedron is an example of transitive but non
Cayley planar graphs. More precisely, there exist infinite families of
graphs having this property.

In this article, we refine the description of the groups of
automorphisms of the graphs and the geometrical properties of their
possible planar embeddings given in \cite{DavidCayley}. We represent
these graphs by their geometrical invariants in a structure called a
{\it labeling scheme}, as long as a special kind of finite state
automaton called a {\it border automaton}. We show that there exists a
bijection between this representation and the class of the TLF-planar
transitive graphs.  Our main result (page~\pageref{thm:enum}) is:

\medskip
\noindent\textbf{Theorem 15 (Enumeration)} \textit{%
Given a number $d\geq 2$, it is possible to enumerate all the 
TLF-planar transitive graphs having internal degree $d$, along
with their labeling schemes.
}
\vskip 1mm

Each vertex-transitive graph belonging to the class $\mathcal{T}$ is
effectively computable ({\it i.e.} there exists an algorithm able to
build every finite ball of the graph). Associated to our results on
Cayley graphs, this allows us to determine which of these graphs are
Cayley, and more precisely:

\medskip
\noindent\textbf{Corollary 16 (Cayley checking)} \textit{%
If $\Gamma$ is a TLF transitive graph, then it is
decidable whether $\Gamma$ is the Cayley graph of a group or not, and
obtain an enumeration and a description of the groups having $\Gamma$
as a Cayley graph.
} 
\vskip 1mm

Finally, thanks to the characterization of the embeddings of the
graphs in $\mathcal{T}$, it is possible to compute their connectivity
and approximate their growth rate, which can be either linear,
quadratic or exponential, depending on their local geometrical
properties. 

\section{TLF-planar transitive graphs}

A \textit{graph} $\Gamma$ consists of a pair $(V,E)$, $V$ being a
countable set of \textit{vertices} and a set of {\it edges}, where $E$
is a subset of the pair of elements of $V$. Each edge corresponds to a
pair of vertices $(v_1,v_2)$ called its extremities. An edge with the
same extremities $(v,v)$ is called a loop. The graphs that we consider
are loopless. An edge $(u,v)$ is said to be \textit{incident} to the
vertices $u$ and $v$.  A {\it labeling} of the graph is an application
from the set of edges into a finite set $L$ of labels or colors. The
{\it degree} of a vertex is the number of edges incident to this
vertex.  A {\it path} of $\Gamma$ is a sequence of vertices $(v_n)$ in
$\Gamma$ such that for all $n$, there exists an edge between $v_n$ and
$v_{n+1}$. A {\it cycle} is a finite path whose initial and terminal
vertices are the same. A \textit{simple} cycle is a cycle where no
vertex appears twice.

\begin{rmk}{Considerations on the construction of subgraphs}
We occasional\-ly build subgraphs of $\Gamma$ by considering a certain
subset of vertices and edges $(V',E')$ where $V'\subset V$ and
$E'\subset E$, or equivalently by removing from the graph a subset of
its vertices and edges. Then, we can consider the remaining set of
vertices and edges as a subspace of the graph seen as a metric space,
and the connected components of this subspace. These components may
not be graphs themselves, since some edges will not have vertices as
their extremities. We can resolve this problem and consider these
components as new graphs by adding new vertices to the extremities of
these edges.
\end{rmk}

A graph $\Gamma$ is {\it connected} if, for every pair of vertices
$(s_1,s_2)$ of the graph, there exists a finite path in the graph with
extremities $s_1$ and $s_2$. A \textit{connected component} is an
equivalence class of vertices for the relation ``to be connected''.
Notice that both definitions are coherent whether $\Gamma$ is seen as
a graph or as a metric space. A {\it n-separation} is a set of $n$
vertices whose removal separates the graph in two or more connected
components not reduced to a single edge. A {\it cut-vertex} of
$\Gamma$ is a $1$-separation of $\Gamma$.  A graph is $n$-{\it
  separable} if it contains a $n$-separation. If it contains no
$n$-separation, it is {\it (n+1)-connected}. A graph is {\it regular}
when all its vertices have the same degree $d$. The graphs we will be
dealing with are connected and regular.

A {\it morphism} from the graph $\Gamma_1 = (V_1,E_1)$ into $\Gamma_2
= (V_2,E_2)$ is an application $\sigma :V_1\rightarrow V_2$ that
preserves the edges of the graph. When both graphs are labeled, we
impose that the morphisms also preserve the labels of the edges. A
graph is said to be {\it vertex-transitive} -- or {\it transitive} in
short -- if and only if, given any two vertices $(s_1,s_2) \in
\Gamma$, there exists an automorphism of $\Gamma$ mapping $s_1$ onto
$s_2$. If $\Gamma$ is the Cayley graph of a group, then it is
vertex-transitive.
 
\begin{rmk}{About the lower degree transitive graphs}
  There exists only one non-trivial connected transitive graph of
  degree $1$, which corresponds to $K_2$, the graph reduced to a
  single edge.  Transitive graphs of degree $2$ correspond to cyclic
  graphs $C_n$ where $n$ may be infinite. These graphs possess
  exactly one labeling when $n$ is odd and two labelings when $n$ is
  even, these labelings corresponding to the planar Cayley graphs of
  degree $2$ associated to the dihedral groups and cyclic groups. In
  the following, we shall only be interested in connected transitive
  graphs of degree $d \geq~3$.
\end{rmk}

A graph $\Gamma$ is said to be {\it planar} if it can be embedded in
the plane, such that no two edges meet in a point other than a common
end. By the plane, we mean a simply connected Riemannian surface,
homogeneous and isotropic. For our embeddings, we will only consider
the three usual geometries : the sphere, the Euclidean and the
hyperbolic plane. Our embeddings will be considered {\it tame},
meaning that all edges are~$\mathcal{C}^1$ images of~$[0;1]$. Such an
embedding is said to be {\it topologically locally finite} -- in short
TLF-planar -- if its vertices have no accumulation point in the
plane. Equivalently, every compact subset of the plane intersects a
finite number of vertices of the embedding. Symmetrically, an
embedding is said to be TLF in terms of edges if and only if every
compact subset of the plane intersects a finite number of edges of the
embedding. The following theorem asserts that a TLF-planar graph
always possess such an embedding:

\begin{thm}[\cite{DavidCayley}] \label{thm:embedding}
  If the graph $\Gamma$ is TLF-planar, there exists a {\it tame}
  embedding of the same graph that is TLF in terms of vertices but also
  of edges.
\end{thm}

We will always suppose that the TLF-planar graphs are embedded in the
plane such that their embedding follows Theorem~\ref{thm:embedding}.
Given a specific embedding of a TLF-planar graph $\Gamma$, a
\textit{face} $\mathcal{F}$ is defined as an arc-connected component
of the complement of the graph in the plane. $\mathcal{F}$ is said to
be \textit{finite} when it is incident to finitely many vertices of
the graph, otherwise it is said to be {\it infinite}. For TLF-planar
graphs, infinite faces are necessarily topologically unbounded in the
plane. The {\it border} of the face $\mathcal{F}$, noted
$\partial\mathcal{F}$, is its boundary in topological terms. A face
$\mathcal{F}$ is said to be \textit{incident} to a vertex or an edge
of the graph if and only if this vertex of edge intersects with
$\partial\mathcal{F}$.

In such an embedding, every edge incident to a face is entirely
included into the border of this face. Then every edge is incident to
exactly two faces of $\Gamma$, which it separates. Considering the
previous definitions of the faces, the transitivity property of
$\Gamma$ stands out with the following lemma taken
from~\cite{DavidCayley}:

\begin{lem}[Intersection of faces]
\label{lem:intersectionfaces}
Let $\Gamma$ be a vertex-transitive TLF-planar graph. Given two
distinct faces of $\Gamma$, the intersection of their border, when
non-empty, is either a vertex or an edge of $\Gamma$.
\end{lem}

\begin{cor}[Preservation of finite faces]
\label{lem:finitefaces}
The automorphisms of $\Gamma$ map the border of every finite face onto
the border of another finite face.
\end{cor}

\begin{rmk}{On the choice of the embedding}
  The previous statements hold for a particular
  embedding of $\Gamma$ that is locally finite in terms of edges and
  vertices. This embedding may not be unique. For example, if $\Gamma$
  is finite, there exists an embedding of $\Gamma$ in the sphere,
  where all faces are topologically bounded. If we select a point
  inside a face and send this point to the infinity, we obtain another
  embedding of the graph on a non-compact surface homeomorphic to the
  Euclidean plane. With the previous definitions, the faces of both
  embeddings are all finite, and the validity of
  Corollary~\ref{lem:finitefaces} is the same for both embeddings.
\end{rmk}

The {\it size} of a face of an embedding of $\Gamma$ corresponds to
the number (possibly infinite) of vertices it is incident to. The {\it
  type vector} of a vertex of $\Gamma$ is the sequence of sizes of the
faces appearing consecutively around this vertex. It is defined up to
rotation and symmetry of the graph. If $\Gamma$ is transitive and
Corollary~\ref{lem:finitefaces} holds, the type vector is independent
of the choice of the vertex, up to permutation of its elements. For
example, the type vector of the Euclidean infinite grid is $[4;4;4;4]$
and the type vector of a cyclic graph $C_n$ with $n$ vertices is
$[n;n]$.

For a given graph $\Gamma$, $\Aut$ denotes its group of automorphisms,
and $\Tran$ stands for the set of subgroups of $\Aut$ acting
transitively on the set of vertices of $\Gamma$. Let $G$ belong to
$\Tran$.  $G$ acts on the set of edges of $\Gamma$ and the set of
orbits of edges is finite. Thus a \textit{class} or a \textit{color}
of edges is defined as an orbit under the action of $G$, the set of
colors being called $\mathsf{E}_G$. In the same manner, we define
classes or colors of finite faces, corresponding to the finite set
$\mathsf{F}_G$. This coloring defines a partition of the set of finite
faces of the embedding. Infinite faces have a special status since
these faces may not be stable by automorphism. Therefore, we request
by convention that all of them correspond to a special color
in~$\mathsf{F}$.

In the following, $\Gamma$ will be a TLF-planar, connected
vertex-transitive graph of finite degree $d\geq 3$. Thus we will speak
of vertices, edges and faces of $\Gamma$, as defined above. $G$ is a
group belonging to $\Tran$. We will always suppose that the embedding
of $\Gamma$ follows Theorem~\ref{thm:embedding} and that the
automorphisms preserve the borders of the finite faces.  In the
section~\ref{sec:localinvariant}, we analyze these local invariants,
in order to obtain a characterization of the graph by its local
geometrical properties in section~\ref{sec:labelingscheme}. The last
section presents some applications of these characterizations.

\section{Local geometrical invariants}
\label{sec:localinvariant}

\subsection{Infinite faces and connectivity}

Let us give some intuition on the general structure of the graphs in
this class. We prove that the finite vertex-transitive planar graphs
of degree $\geq 3$ are all $3$-connected graphs, and in the infinite
case, the connectivity of the graphs depends only on the number of
infinite faces appearing around each vertex:

\begin{lem}[Connectivity and infinite faces] \label{lem:connectivity}
  If $\Gamma$ is a TLF-planar transitive graph of degree $\geq 3$, let
  $n$ be the number of infinite faces appearing around a given vertex
  of $\Gamma$. Then, depending on the value of $n$: \par\nopagebreak
\noindent {\parbox{5.4cm}{\begin{itemize}
\setlength{\topsep}{0pt}%
\setlength{\itemsep}{0pt}%
\setlength{\parskip}{0pt}%
\item $n\geq2 \Leftrightarrow \Gamma$ is 1-separable.
\item $n=1 \Leftrightarrow \Gamma$ is 2-connected and 2-separable;
\item $n=0 \Leftrightarrow \Gamma$ is 3-connected;
\end{itemize}}} \hfill 
{\parbox{8.0cm}{{\begin{pspicture}(-2.1,-1.2)(0,0.8)
\psset{unit=0.7}

\cnode(-1.5,0){2pt}{A}
\cnode(-1.7,0.5){2pt}{A1}       \ncline{A}{A1}
\cnode(-2,0){2pt}{A2}           \ncline{A}{A2}
\cnode(-1.6,-0.6){2pt}{A3}      \ncline{A}{A3}
\cnode(-1.1,-0.3){2pt}{A4}      \ncline{A}{A4}
\cnode(-1,0.4){2pt}{A5}         \ncline{A}{A5}

\pscurve[linestyle=dashed,linecolor=gray](-1.7,0.5)(-2.5,0.8)(-2,0)
\pscurve[linestyle=dashed,linecolor=gray](-2,0)(-2.5,-0.8)(-1.6,-0.6)
\pscurve[linestyle=dashed,linecolor=gray](-1.6,-0.6)(-1,-1.1)(-1.1,-0.3)
\pscurve[linestyle=dashed,linecolor=gray](-1.1,-0.3)(-0.6,0)(-1,0.4)
\pscurve[linestyle=dashed,linecolor=gray](-1,0.4)(-1.2,1)(-1.7,0.5)

\rput[c](-1.5,-1.5){$n=0$}

\end{pspicture}}
{\begin{pspicture}(-2.8,-1.2)(0,0.8)
\psset{unit=0.7}

\cnode(-1.5,0){2pt}{A}
\cnode(-1.7,0.5){2pt}{A1}       \ncline{A}{A1}
\cnode(-2.5,0){2pt}{A2}         \ncline{A}{A2}
\cnode(-1.6,-0.6){2pt}{A3}      \ncline{A}{A3}
\cnode(-1.1,-0.3){2pt}{A4}      \ncline{A}{A4}
\cnode(-1,0.4){2pt}{A5}         \ncline{A}{A5}

\cnode(-2.3,0.5){2pt}{B1}       \ncline{A2}{B1}
\cnode(-2.4,-0.6){2pt}{B3}      \ncline{A2}{B3}
\cnode(-2.9,-0.3){2pt}{B4}      \ncline{A2}{B4}
\cnode(-3,0.4){2pt}{B5}         \ncline{A2}{B5}

\pscurve[linestyle=dashed,linecolor=gray](-1.6,-0.6)(-1,-1.1)(-1.1,-0.3)
\psline[linewidth=2pt,linecolor=lightred,linestyle=dotted](-1.2,0)(0,0)
\pscurve[linestyle=dashed,linecolor=gray](-1,0.4)(-1.2,1)(-1.7,0.5)
\pscurve[linestyle=dashed,linecolor=gray](-2.4,-0.6)(-3,-1.1)(-2.9,-0.3)
\psline[linewidth=2pt,linecolor=lightred,linestyle=dotted](-2.8,0)(-4,0)
\pscurve[linestyle=dashed,linecolor=gray](-3,0.4)(-2.8,1)(-2.3,0.5)
\nccurve[linestyle=dashed,linecolor=gray,angleA=45,angleB=135]{B1}{A1}
\nccurve[linestyle=dashed,linecolor=gray,angleA=315,angleB=225]{B3}{A3}

\rput[c](-2,-1.5){$n=1$}

\end{pspicture}}
{\begin{pspicture}(-2.1,-1.2)(0,0.8)
\psset{unit=0.7}

\cnode(-1.5,0){2pt}{A}
\cnode(-1.7,0.5){2pt}{A1}       \ncline{A}{A1}
\cnode(-2,0){2pt}{A2}           \ncline{A}{A2}
\cnode(-1.6,-0.6){2pt}{A3}      \ncline{A}{A3}
\cnode(-1.1,-0.3){2pt}{A4}      \ncline{A}{A4}
\cnode(-1,0.4){2pt}{A5}         \ncline{A}{A5}

\pscurve[linestyle=dashed,linecolor=gray](-1.7,0.5)(-2.5,0.8)(-2,0)
\psline[linewidth=2pt,linecolor=lightred,linestyle=dotted](-1.2,0)(0,0)
\pscurve[linestyle=dashed,linecolor=gray](-1.6,-0.6)(-1,-1.1)(-1.1,-0.3)
\psline[linewidth=2pt,linecolor=lightred,linestyle=dotted](-1.7,-0.2)(-2.5,-1)
\pscurve[linestyle=dashed,linecolor=gray](-1,0.4)(-1.2,1)(-1.7,0.5)

\rput[c](-1.5,-1.5){$n\geq 2$}

\end{pspicture}}}}
\end{lem}

\begin{prf}
\begin{itemize}
\item({\itshape $n\geq 2 \Leftrightarrow\Gamma$ is $1$-separable})\\
  Given a
  vertex $v$ of $\Gamma$, we consider the set of faces incident to
  that vertex, and the union of the borders of those faces that are
  finite. If $n\geq 2$, the union of $v$ and two infinite faces
  incident to $v$ separates the graph into at least two non-trivial
  components. Then every vertex of the graph is a cut-vertex and the
  graph is $1$-separable. On the other hand, if $\Gamma$ is
  $1$-separable, every vertex must meet at least $2$ infinite faces.

\item({\itshape $n= 1 \Rightarrow\Gamma$ is $2$-connected and
 $2$-separable})\\ 
  If $n=1$, then consider an edge of $\Gamma$ that
 does not belong to the border of an infinite face. There must exist
 one, otherwise $\Gamma$ being of degree at least $3$, that would
 contradict the fact that $n=1$. The extremities of this edge both
 meet an infinite face, and these faces are distinct. The removal of
 these extremities separates $\Gamma$, therefore $\Gamma$ is
 2-separable. It is $2$-connected because~$n<2$.

\item({\itshape $\Gamma$ is $2$-connected and $2$-separable
  $\Rightarrow n=1$})\\ %
  Suppose now that $\Gamma$ is $2$-separable
  and $2$-connected, and consider $\{s,t\}$ a $2$-separation of
  $\Gamma$. Let~$\Lambda$ be a subgraph separated by
  $\{s,t\}$. Suppose that we remove $\Lambda$ from the embedding. The
  remaining TLF-planar graph possesses a face $\mathcal{F}$, inside
  which $\Lambda$ was embedded. Moreover, $s$ and $t$ both belong to
  the border of $\mathcal{F}$. If $\mathcal{F}$ is finite, embedding
  $\Lambda$ inside $\mathcal{F}$ separates the face into at least two
  subfaces meeting at $s$ and $t$, therefore contradicting
  Lemma~\ref{lem:intersectionfaces}. Therefore $\mathcal{F}$ is
  infinite. When embedding $\Lambda$ inside $\mathcal{F}$, there will
  remain an infinite face in the embedding of $\Gamma$. Therefore
  $n\geq 1$ and since $\Gamma$ is $2$-separable, $n=1$.
\end{itemize}
\end{prf}

If $\Gamma$ is 1-separable, then every vertex is a cut-vertex. If we
cut the graph along its cut-vertices, the remaining components are
{\it 2-connected components}. Since the graph is vertex-transitive,
the set of components incident to a vertex is independent of the
vertex, and finite, because the degree of the graph is finite. These
components are TLF-planar graphs, but not necessarily
vertex-transitive themselves. They may be finite or infinite. They may
be reduced to a single edge. If $\Gamma$ is at least 2-connected, it
is composed of a unique 2-connected component equal to $\Gamma$.

\subsection{A simple invariant}

Consider more closely the implications of
Corollary~\ref{lem:finitefaces}. The group $G$ of automorphisms of the
graph acts on the set of the finite faces of the embedding. Let us
focus on the invariants under this action. The classes or colors of
the edges and faces are simple examples of geometrical invariants.

For the sake of clarity, we always mark classes (or colors) of edges
with gothic letters $\mathsf{E}_G=\{\mathfrak{a}, \mathfrak{b},
\mathfrak{c} \dots\}$ and classes (or colors) of faces with greek
letters $\mathsf{F}_G=\{ \alpha, \beta, \gamma \dots\}$. In the
remaining of the article, we will suppose that the group $G$ is fixed
and therefore drop the letter $G$.

\noindent
\parbox{\textwidth}{\begin{lem}[Face separation]
\label{lem:separation}
Consider an edge $e$ of $\Gamma$ belonging to the class
$\mathfrak{e}\in\mathsf{E}$ of edges under the action of $G$ . Then
the classes of faces separated by $e$ are the same independently of
the representative $e\in\mathfrak{e}$.
\end{lem}}

\begin{prf}
Suppose $e$ is mapped by automorphism on $f\in \mathfrak{e}$. As a
result from Theorem~\ref{lem:finitefaces}, the finite faces incident
to the edge $e$ are mapped by automorphism onto the finite faces
incident to $f$. And the automorphism is invertible, therefore the
finite faces incident to both edges are in bijection. In turn, there
is an equal number of infinite faces incident to both edges.  This
concludes the proof.
\end{prf}

According to Lemma~\ref{lem:separation}, for any class of edges
$\mathfrak{e} \in \mathsf{E}$, it is possible to define the {\it
  separator} of $\mathfrak{e}$, namely $\mathsf{sep}(\mathfrak{e})$,
as the pair of classes of faces separated by~$\mathfrak{e}$.
This separator is a geometrical invariant under the action of $G$.

\subsection{Edge and Face vectors}

Given a vertex $v\in\Gamma$, consider the finite subgraph $\Lambda$ of
$\Gamma$ composed of all edges incident to $v$ and its planar
embedding induced by the embedding of~$\Gamma$. Select a particular
edge $e$ incident to $v$. An {\it edge vector} $\xi$ of $\Gamma$
around $v$ is the vector whose elements describe the classes of edges
appearing around $v$ in $\Lambda$ in the positive direction, starting
from $e$.  Similarly, a {\it face vector} $\phi$ around $v$ is the
vector whose elements describe the classes of faces appearing around
$v$ in $\Lambda$, starting from the face next to $e$ in the positive
direction.  As a convention, we always choose the pair $(\xi,\phi)$
composed of an edge vector and a face vector, to be {\it locked} as
follows: the edge $\xi_i$ separates the faces $\phi_i$ and
$\phi_{i-1}$.
We decompose these vectors into blocks separated by infinite faces~:

\begin{center}
  \begin{tabular}{c@{\hskip 1pt}c@{\hskip 1pt}c@{\hskip 1pt}c@{\hskip 1pt}c}
    $\xi$ : [& 
      $\overbrace{\xi_{k_1^s},\dots,\xi_{k_1^e-1},\xi_{k_1^e}}^{\textrm{Block 1}}$, 
      &
      $\overbrace{\xi_{k_2^s},\dots,\xi_{k_2^e-1},\xi_{k_2^e}}^{\textrm{Block 2}}$, 
      & \dots, &
      $\overbrace{\xi_{k_t^s},\dots,\xi_{k_t^e-1},\xi_{k_t^e}}^{\textrm{Block t}}$] 
    \\
    $\phi$ : [& 
      $\underbrace{\phi_{k_1^s},\dots,\phi_{k_1^e-1}}_{\textrm{Block 1}},\infty$,
      &
      $\underbrace{\phi_{k_2^s},\dots,\phi_{k_2^e-1}}_{\textrm{Block 2}},\infty$,
      & \dots, &
      $\underbrace{\phi_{k_t^s},\dots,\phi_{k_t^e-1}}_{\textrm{Block t}},\infty$]
  \end{tabular}
\end{center}
  
Here $\xi_k$ and $\phi_k$ represent respectively the $k$-th elements
of the vectors $\xi$ and $\phi$. The $i$-th block starts at index
$k_i^s$ and ends at index $k_i^e$. If the graph does not contain any
infinite face, then the decomposition contains a single block.

Consider the set of all possible edge and face vectors in the
embedding of $\Gamma$. Define the following operations on this set:

\begin{itemize}
\item \textsc{Rotation and Symmetry:} These operations correspond to
the usual isometries of the plane acting on the face and edge vectors.


\item \textsc{Rearrangement:} The operations consists in rearranging
  the blocks while preserving the fact that these blocks are separated
  by infinite faces. More graphically, given a permutation $\sigma$ of
  the blocks~:

\vskip 3pt
\begin{center}
    \begin{tabular}{c@{\hskip 0pt}c@{\hskip 0pt}c@{\hskip 0pt}c@{\hskip 0pt}c}
      $\xi~$: [& 
      $\overbrace{\xi_{k_{\sigma(1)}^s},\dots,
        \xi_{k_{\sigma(1)}^e}}^{\textrm{Block $\sigma(1)$}}$, 
      &
      $\overbrace{\xi_{k_{\sigma(2)}^s},\dots,
        \xi_{k_{\sigma(2)}^e}}^{\textrm{Block $\sigma(2)$}}$, 
      & \dots, &
      $\overbrace{\xi_{k_{\sigma(t)}^s},\dots,
        \xi_{k_{\sigma(t)}^e}}^{\textrm{Block $\sigma(t)$}}$] 
\\
      $\phi~$: [& 
      $\underbrace{\phi_{k_{\sigma(1)}^s},\dots,
        \phi_{k_{\sigma(1)}^e-1}}_{\textrm{Block $\sigma(1)$}},\infty$,
      &
      $\underbrace{\phi_{k_{\sigma(2)}^s},\dots,
        \phi_{k_{\sigma(1)}^e-1}}_{\textrm{Block $\sigma(2)$}},\infty$,
      & \dots, &
      $\underbrace{\phi_{k_{\sigma(t)}^s},\dots,
        \phi_{k_{\sigma(1)}^e-1}}_{\textrm{Block $\sigma(t)$}},\infty$]
      \end{tabular}
\end{center}
\vskip 3pt

\item \textsc{Twist:} The twist operation describes a symmetry applied
  on a single 2-connected component around a vertex. For example, the
  twist applied onto the first component corresponds to the following
  transformation:

\vskip 5pt
\begin{center}
    \begin{tabular}{c@{\hskip 1pt}c@{\hskip 1pt}c@{\hskip 1pt}c@{\hskip 1pt}c}
      $\xi$ : [& 
      $\overbrace{\xi_{k_1^e},\dots,\xi_{k_1^s+1},\xi_{k_1^s}}^{\textrm{Block
      1 reversed}}$, 
      &
      $\overbrace{\xi_{k_2^s},\dots,\xi_{k_2^e-1},\xi_{k_2^e}}^{\textrm{Block 2}}$, 
      & \dots, &
      $\overbrace{\xi_{k_t^s},\dots,\xi_{k_t^e-1},\xi_{k_t^e}}^{\textrm{Block t}}$] 
\\
      $\phi$ : [& 
      $\underbrace{\phi_{k_1^e-1},\dots,\phi_{k_1^s}}_{\textrm{Block 1 reversed}},\infty$,
      &
      $\underbrace{\phi_{k_2^s},\dots,\phi_{k_2^e-1}}_{\textrm{Block 2}},\infty$,
      & \dots, &
      $\underbrace{\phi_{k_t^s},\dots,\phi_{k_t^e-1}}_{\textrm{Block t}},\infty$]
      \end{tabular}
\end{center}
\end{itemize}

Two pairs of locked vectors $(\xi_1,\phi_i)$ and $(\xi_2,\phi_2)$
around $v$ are said to be {\it isomorphic} if and only if it is
possible to transform the first into the second by a sequence of
rotations, symmetries, rearrangements and twists. The following lemma
states that these operations describe all the possible edge and face
vectors in the embedding:

\begin{lem}[Edge and Face vectors] \label{lem:edgefacevect}
The edge vector and the face vector of $\Gamma$ is independent of the
choice of the embedding of $\Gamma$ and of the vertex around which it
is chosen, up to isomorphism.
\end{lem}

\begin{prf}
  This is a direct consequence of Corollary~\ref{lem:finitefaces}.
  Since finite faces are mapped onto finite faces by the automorphisms
  of $\Gamma$, then the 2-connected components of $\Gamma$ are mapped
  onto 2-connected components. Therefore, the only operations that we
  may apply on the set of edge and face vectors of $\Gamma$ are
  rotations and symmetries in the case of 2-connected graphs, and
  rearrangement of the $2$-connected components for 1-separable
  graphs. These operations are exactly those described by the
  twists and rearrangements. 
\end{prf}

Given an edge vector $\xi$ and a face vector $\phi$ of $\Gamma$ that
are locked together, it is therefore possible to determine all
possible vectors in the class of isomorphism. Therefore we will only
consider {\it the} edge and face vectors of $\Gamma$ by choosing a
representative in this class.  

\begin{exm} \label{exm:edgefacevector}
Suppose that the graph $\Gamma$ possesses the set of colors defined by
  $\mathsf{E}=\{\mathfrak{b},\mathfrak{r},\mathfrak{g}\}$ for its
  edges and $\mathsf{F}=\{\alpha,\beta,\gamma\}$ for its faces. An
  example of such a graph appears on Figure~\ref{fig:example2}
  page~\pageref{fig:example2}. We represent the edge and face vectors
  $(\xi,\phi)$ of this graph by the following picture~:
  
  \begin{center}
    \begin{pspicture}(-2.5,-2.4)(2.5,2.4)
      \SpecialCoor
      \degrees[5] \psset{unit=1.35}
      
      \pscustom[fillstyle=solid,fillcolor=lightora,linecolor=orange,linearc=0.25]{
        \pspolygon(0.1;0.5)(1.2;0.1)(1.2;0.9)}
      \rput[c](0.725;0.5){$\textcolor{orange}{\beta}$}
      \pscustom[fillstyle=solid,fillcolor=lightvio,linecolor=violet,linearc=0.25]{
        \pspolygon(0.1;1.5)(1.2;1.1)(1.2;1.9)}
      \rput[c](0.725;1.5){$\textcolor{violet}{\gamma}$}
      \pscustom[fillstyle=solid,fillcolor=lightgreen,linecolor=green,linearc=0.25]{
        \pspolygon(0.1;2.5)(1.2;2.1)(1.2;2.9)}
      \rput[c](0.725;2.5){$\textcolor{green}{\alpha}$}
      \pscustom[fillstyle=solid,fillcolor=lightora,linecolor=orange,linearc=0.25]{
        \pspolygon(0.1;3.5)(1.2;3.1)(1.2;3.9)}
      \rput[c](0.725;3.5){$\textcolor{orange}{\beta}$}
      \pscustom[fillstyle=solid,fillcolor=lightvio,linecolor=violet,linearc=0.25]{
        \pspolygon(0.1;4.5)(1.2;4.1)(1.2;4.9)}
      \rput[c](0.725;4.5){$\textcolor{violet}{\gamma}$}

      \cnode(0,0){2pt}{A}
      \cnode(1,0){2pt}{A1}            \ncline[linecolor=lightvio]{A}{A1}
      \rput[c](1.30;0){$1(\textcolor{red}{\mathfrak{r}})$}
      \cnode(0.3,0.9){2pt}{A2}        \ncline[linecolor=lightvio]{A}{A2}
      \rput[c](1.25;1){$2(\textcolor{red}{\mathfrak{r}})$}
      \cnode(-0.8,0.6){2pt}{A3}       \ncline[linecolor=black]{A}{A3}
      \rput[c](1.30;2){$3(\textcolor{black}{\mathfrak{b}})$}
      \cnode(-0.8,-0.6){2pt}{A4}      \ncline[linecolor=notwhite]{A}{A4}
      \rput[c](1.30;3){$4(\textcolor{gold}{\mathfrak{g}})$}
      \cnode(0.3,-0.9){2pt}{A5}       \ncline[linecolor=lightvio]{A}{A5}
      \rput[c](1.25;4){$5(\textcolor{red}{\mathfrak{r}})$}

      \psarc{->}(0,0){0.25}{0}{4}

    \end{pspicture}
    ~
    \raise 2.3cm \hbox{\parbox{6cm}{%
      \begin{tabular}{c@{\hskip 4pt}cc}
        Edge vector & $\xi~:$ &
        $[\mathfrak{r},\mathfrak{r},\mathfrak{b},\mathfrak{g},\mathfrak{r}]$
        \\ 
        Face vector & $\phi~:$ &
        $[\beta,\gamma,\alpha,\beta,\gamma]$ \\
      \end{tabular}}}
  \end{center}

Suppose for our example that the faces colored by $\beta$ are
infinite. Therefore, in the aforementioned decomposition, there are
two blocks, one containing the edges numbered $\{1;5\}$ and the other
containing the edges numbered $\{2;3;4\}$. Under these hypotheses,
we can operate the following transformations onto the pair $(\xi,\phi)$:

{~\hfill
    {\begin{pspicture}(-1.5,-2)(1.5,1.5)
      \SpecialCoor
      \degrees[5] \psset{unit=0.85}
      
      \pscustom[fillstyle=solid,fillcolor=lightora,linecolor=orange,linearc=0.25]{
        \pspolygon(0.1;1.5)(1.2;1.1)(1.2;1.9)}
      \rput[c](0.725;1.5){\scaleboxto(0,0.4){$\textcolor{orange}{\strut\beta}$}}
      \pscustom[fillstyle=solid,fillcolor=lightvio,linecolor=violet,linearc=0.25]{
        \pspolygon(0.1;2.5)(1.2;2.1)(1.2;2.9)}
      \rput[c](0.725;2.5){\scaleboxto(0,0.4){$\textcolor{violet}{\strut\gamma}$}}
      \pscustom[fillstyle=solid,fillcolor=lightgreen,linecolor=green,linearc=0.25]{
        \pspolygon(0.1;3.5)(1.2;3.1)(1.2;3.9)}
      \rput[c](0.725;3.5){\scaleboxto(0,0.4){$\textcolor{green}{\strut\alpha}$}}
      \pscustom[fillstyle=solid,fillcolor=lightora,linecolor=orange,linearc=0.25]{
        \pspolygon(0.1;4.5)(1.2;4.1)(1.2;4.9)}
      \rput[c](0.725;4.5){\scaleboxto(0,0.4){$\textcolor{orange}{\strut\beta}$}}
      \pscustom[fillstyle=solid,fillcolor=lightvio,linecolor=violet,linearc=0.25]{
        \pspolygon(0.1;0.5)(1.2;0.1)(1.2;0.9)}
      \rput[c](0.725;0.5){\scaleboxto(0,0.4){$\textcolor{violet}{\strut\gamma}$}}

      \cnode(0,0){2pt}{A}
      \cnode(1;1){2pt}{A1}            \ncline[linecolor=lightvio]{A}{A1}
      \rput[c](1.50;1){\scaleboxto(0,0.4){$2(\textcolor{red}{\mathfrak{r}})$}}
      \cnode(1;2){2pt}{A2}        \ncline[linecolor=lightvio]{A}{A2}
      \rput[c](1.45;2){\scaleboxto(0,0.4){$3(\textcolor{red}{\mathfrak{r}})$}}
      \cnode(1;3){2pt}{A3}       \ncline[linecolor=black]{A}{A3}
      \rput[c](1.50;3){\scaleboxto(0,0.4){$4(\textcolor{black}{\mathfrak{b}})$}}
      \cnode(1;4){2pt}{A4}      \ncline[linecolor=notwhite]{A}{A4}
      \rput[c](1.50;4){\scaleboxto(0,0.4){$5(\textcolor{gold}{\mathfrak{g}})$}}
      \cnode(1;0){2pt}{A5}       \ncline[linecolor=lightvio]{A}{A5}
      \rput[c](1.45;0){\scaleboxto(0,0.4){$1(\textcolor{red}{\mathfrak{r}})$}}

      \psarc{->}(0,0){0.25}{0}{4}

      \rput(0,-2){Rotation}
    \end{pspicture}}
    \hfill
    {\begin{pspicture}(-1.5,-2)(1.5,1.5)
      \SpecialCoor
      \degrees[5] \psset{unit=0.85}
      
      \pscustom[fillstyle=solid,fillcolor=lightora,linecolor=orange,linearc=0.25]{
        \pspolygon(0.1;4.5)(1.2;4.1)(1.2;4.9)}
      \rput[c](0.725;4.5){\scaleboxto(0,0.4){$\textcolor{orange}{\strut\beta}$}}
      \pscustom[fillstyle=solid,fillcolor=lightvio,linecolor=violet,linearc=0.25]{
        \pspolygon(0.1;3.5)(1.2;3.1)(1.2;3.9)}
      \rput[c](0.725;3.5){\scaleboxto(0,0.4){$\textcolor{violet}{\strut\gamma}$}}
      \pscustom[fillstyle=solid,fillcolor=lightgreen,linecolor=green,linearc=0.25]{
        \pspolygon(0.1;2.5)(1.2;2.1)(1.2;2.9)}
      \rput[c](0.725;2.5){\scaleboxto(0,0.4){$\textcolor{green}{\strut\alpha}$}}
      \pscustom[fillstyle=solid,fillcolor=lightora,linecolor=orange,linearc=0.25]{
        \pspolygon(0.1;1.5)(1.2;1.1)(1.2;1.9)}
      \rput[c](0.725;1.5){\scaleboxto(0,0.4){$\textcolor{orange}{\strut\beta}$}}
      \pscustom[fillstyle=solid,fillcolor=lightvio,linecolor=violet,linearc=0.25]{
        \pspolygon(0.1;0.5)(1.2;0.1)(1.2;0.9)}
      \rput[c](0.725;0.5){\scaleboxto(0,0.4){$\textcolor{violet}{\strut\gamma}$}}

      \cnode(0,0){2pt}{A}
      \cnode(1;0){2pt}{A1}            \ncline[linecolor=lightvio]{A}{A1}
      \rput[c](1.50;0){\scaleboxto(0,0.4){$1(\textcolor{red}{\mathfrak{r}})$}}
      \cnode(1;4){2pt}{A2}        \ncline[linecolor=lightvio]{A}{A2}
      \rput[c](1.45;4){\scaleboxto(0,0.4){$5(\textcolor{red}{\mathfrak{r}})$}}
      \cnode(1;3){2pt}{A3}       \ncline[linecolor=black]{A}{A3}
      \rput[c](1.50;3){\scaleboxto(0,0.4){$4(\textcolor{black}{\mathfrak{b}})$}}
      \cnode(1;2){2pt}{A4}      \ncline[linecolor=notwhite]{A}{A4}
      \rput[c](1.50;2){\scaleboxto(0,0.4){$3(\textcolor{gold}{\mathfrak{g}})$}}
      \cnode(1;1){2pt}{A5}       \ncline[linecolor=lightvio]{A}{A5}
      \rput[c](1.45;1){\scaleboxto(0,0.4){$2(\textcolor{red}{\mathfrak{r}})$}}

      \psarc{->}(0,0){0.25}{0}{4}

      \rput(0,-2){Symmetry}
    \end{pspicture}}
    \hfill
    {\begin{pspicture}(-1.5,-2)(1.5,1.5)
      \SpecialCoor
      \degrees[5] \psset{unit=0.85}
      
      \pscustom[fillstyle=solid,fillcolor=lightora,linecolor=orange,linearc=0.25]{
        \pspolygon(0.1;0.5)(1.2;0.1)(1.2;0.9)}
      \rput[c](0.725;0.5){\scaleboxto(0,0.4){$\textcolor{orange}{\strut\beta}$}}
      \pscustom[fillstyle=solid,fillcolor=lightvio,linecolor=violet,linearc=0.25]{
        \pspolygon(0.1;2.5)(1.2;2.1)(1.2;2.9)}
      \rput[c](0.725;2.5){\scaleboxto(0,0.4){$\textcolor{violet}{\strut\gamma}$}}
      \pscustom[fillstyle=solid,fillcolor=lightgreen,linecolor=green,linearc=0.25]{
        \pspolygon(0.1;1.5)(1.2;1.1)(1.2;1.9)}
      \rput[c](0.725;1.5){\scaleboxto(0,0.4){$\textcolor{green}{\strut\alpha}$}}
      \pscustom[fillstyle=solid,fillcolor=lightora,linecolor=orange,linearc=0.25]{
        \pspolygon(0.1;3.5)(1.2;3.1)(1.2;3.9)}
      \rput[c](0.725;3.5){\scaleboxto(0,0.4){$\textcolor{orange}{\strut\beta}$}}
      \pscustom[fillstyle=solid,fillcolor=lightvio,linecolor=violet,linearc=0.25]{
        \pspolygon(0.1;4.5)(1.2;4.1)(1.2;4.9)}
      \rput[c](0.725;4.5){\scaleboxto(0,0.4){$\textcolor{violet}{\strut\gamma}$}}

      \cnode(0,0){2pt}{A}
      \cnode(1;0){2pt}{A1}            \ncline[linecolor=lightvio]{A}{A1}
      \rput[c](1.50;0){\scaleboxto(0,0.4){$1(\textcolor{red}{\mathfrak{r}})$}}
      \cnode(1;3){2pt}{A2}        \ncline[linecolor=lightvio]{A}{A2}
      \rput[c](1.45;3){\scaleboxto(0,0.4){$4(\textcolor{red}{\mathfrak{r}})$}}
      \cnode(1;2){2pt}{A3}       \ncline[linecolor=black]{A}{A3}
      \rput[c](1.50;2){\scaleboxto(0,0.4){$3(\textcolor{black}{\mathfrak{b}})$}}
      \cnode(1;1){2pt}{A4}      \ncline[linecolor=notwhite]{A}{A4}
      \rput[c](1.50;1){\scaleboxto(0,0.4){$2(\textcolor{gold}{\mathfrak{g}})$}}
      \cnode(1;4){2pt}{A5}       \ncline[linecolor=lightvio]{A}{A5}
      \rput[c](1.45;4){\scaleboxto(0,0.4){$5(\textcolor{red}{\mathfrak{r}})$}}

      \psarc{->}(0,0){0.25}{0}{4}

      \rput(0,-2){Twist}
    \end{pspicture}}
\hfill~}

With only two blocks, a rearrangement is the same as a
rotation. Moreover, since one of the blocks is stable by symmetry, a
twist of this component leaves the pair unchanged. A twist of the
other component is the same as a symmetry of the pair. 
\end{exm}

\begin{exm}
 Edge vectors and face vectors in general do not determine a
vertex-transitive graph in a unique way. It is quite possible to
obtain non-isomorphic graphs possessing the same edge and face
vector. For example, consider the graphs on Figure~\ref{fig:edgeface}.
Both graphs correspond to the planar tiling of the hyperbolic plane
with decagons; both are vertex-transitive and face-transitive graphs,
and they have the same edge vectors. Nevertheless, the borders of the
faces differ: for the graph on the left, it corresponds to
$(\mathfrak{rgrbrbgbrb})$ and $(\mathfrak{rgbrb})^2$ on the right,
where $\mathfrak{r}$, $\mathfrak{g}$ and $\mathfrak{b}$ respectively
stand for the three different classes of edges.
\begin{figure}[ht] 
\begin{center}
\begin{pspicture}(-2,-2.5)(2,2.6)
\rput(-4,0){\includegraphics[width=6cm]{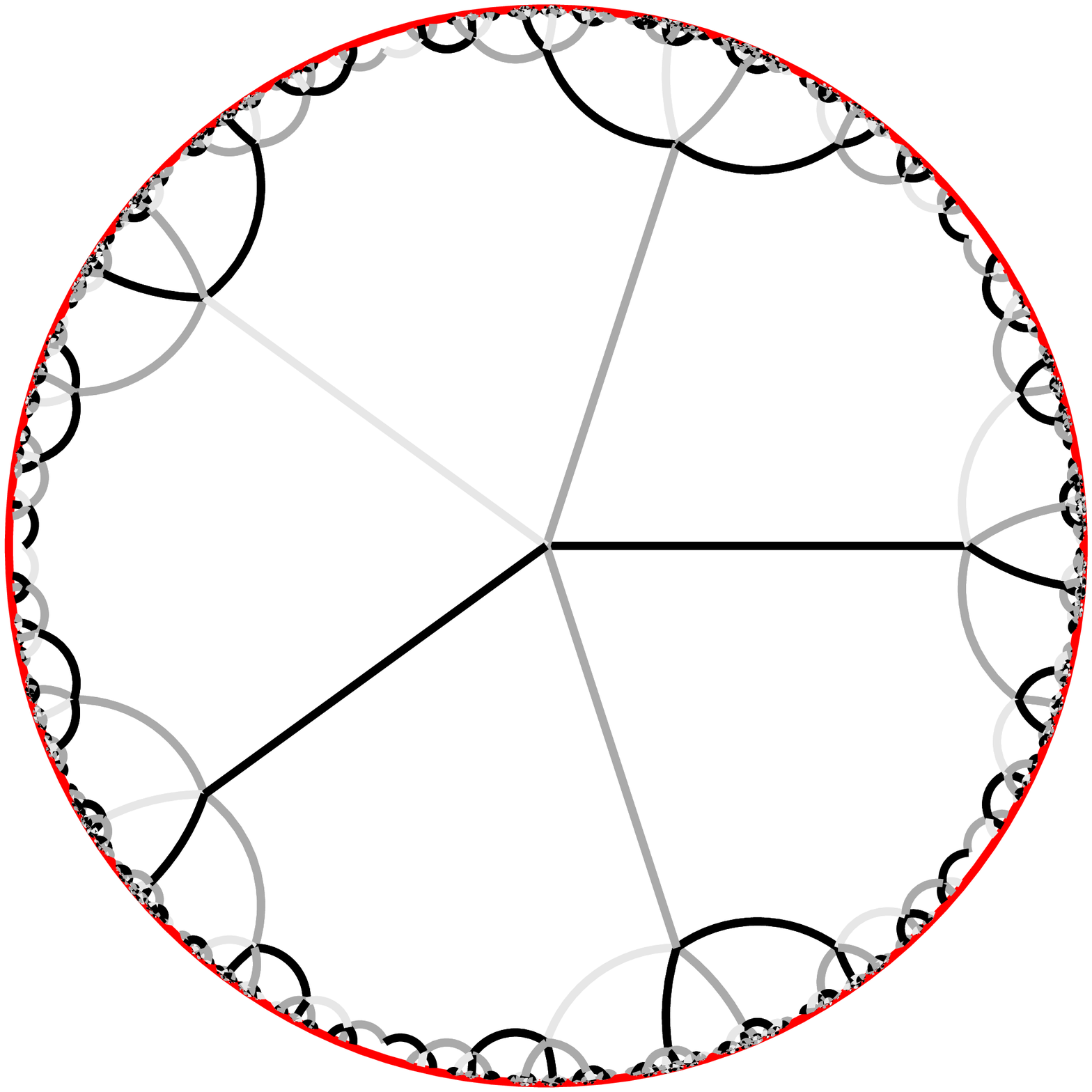}}
\psframe[linecolor=red,linestyle=dashed](-6.6,0.6)(-5,2.2)
\psclip{
\psframe[linestyle=none](-1,-0.4)(1,-2.4)
}
\rput(2.9,-3.6){\includegraphics[width=10cm]{figure1.ps}}
\endpsclip
\psframe[linecolor=lightgray,linestyle=dashed,linewidth=0.2pt](-1,-0.4)(1,-2.4)

\rput(4.2,0){\includegraphics[width=6cm]{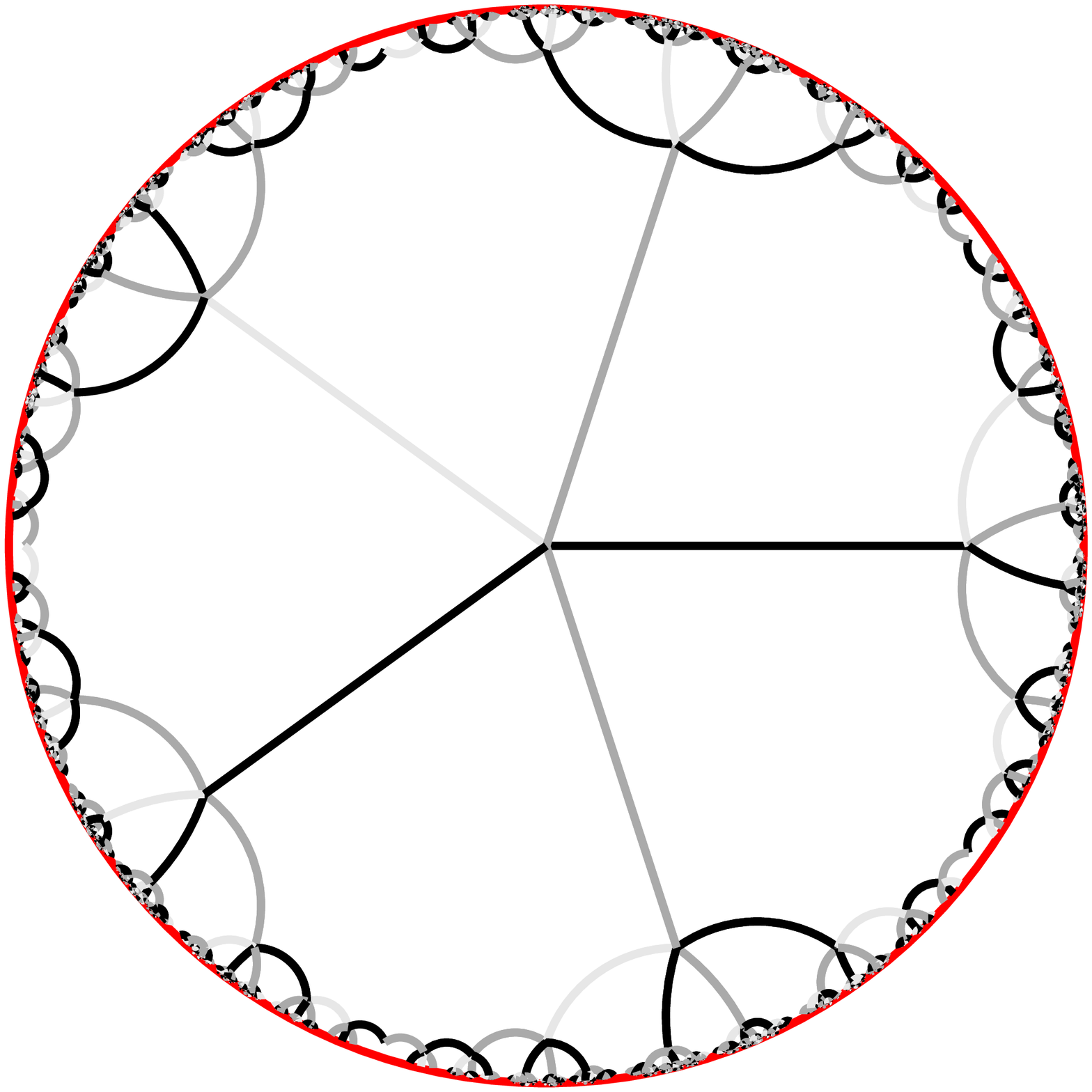}}
\psframe[linecolor=red,linestyle=dashed](1.6,0.6)(3.2,2.2)
\psclip{
\psframe[linestyle=none](-1,0.4)(1,2.4)
}
\rput(2.9,-0.8){\includegraphics[width=10cm]{figure2.ps}}
\endpsclip
\psframe[linecolor=lightgray,linestyle=dashed,linewidth=0.2pt](-1,0.4)(1,2.4)

\psline[linestyle=dashed,linecolor=red](1.6,1.4)(1,1.4)
\psline[linestyle=dashed,linecolor=red](-5,0.6)(-3,-1.4)(-1,-1.4)

\end{pspicture}
\end{center}
\caption{Non-isomorphic transitive graphs with the same edge
  and type vectors. The extremities of the lighter edge are
  non-isomorphic (see detail), which, considering the fact that both
  graphs are 3-connected, implies the non-isomorphism.}
\label{fig:edgeface}
\end{figure}
\end{exm}

For an accurate description of the graph, some complementary
informations are therefore needed. Following the intuitions in the
case of TLF-planar Cayley graphs \cite{DavidCayley}, these informations are
likely to come from local invariants linked to the classes of edges of
$\Gamma$.

\subsection{Edge neighborhoods}

\label{sec:edgeneighb}
\parbox{10.3cm}{\parfillskip=0pt Let $e$ be an edge of $\Gamma$, and
$T$ the finite tree composed of the edge $e$ and all edges incident to
both extremities of $e$. An {\it edge neighborhood} for $e$ is a
structure composed of the two edge vectors representing the classes of
edges appearing around each extremity of $e$ in the same direction,
where the first element of each vector corresponds to $e$, and two
face vectors representing the classes of faces appearing around each }
~ \lower 1.4cm 
\hbox{{{\begin{pspicture}(0.1,1)(3.2,4.4)
\psset{xunit=1.7,yunit=1.7}
\pscustom[linestyle=none,fillstyle=solid,fillcolor=notwhite]{%
\psline(0.25,1)(1,1) \psline(1,1)(1,2) \psline(1,2)(0.25,2)}
\rput(0.6,1.5){$\phi_1$}
\pscustom[linestyle=none,fillstyle=solid,fillcolor=notwhite]{%
\psline(1.75,1)(1,1) \psline(1,1)(1,2) \psline(1,2)(1.75,2)}
\rput(1.4,1.5){$\phi'_1$}

\cnode(1,1){2.5pt}{A}
\cnode(1,2){2.5pt}{B}     \ncline[linecolor=turquoise,linewidth=2pt]{A}{B}
\cnode(0.5,1){2.5pt}{A1}  \ncline[linecolor=green,linewidth=2pt]{A}{A1}
\cnode(1.5,1){2.5pt}{A2}
\psdots[linecolor=green,dotscale=0.6](1.09,1)
\psdots[linecolor=green,dotscale=0.6](1.17,1)
\psdots[linecolor=green,dotscale=0.6](1.25,1)
\psdots[linecolor=green,dotscale=0.6](1.33,1)
\psdots[linecolor=green,dotscale=0.6](1.41,1)

\cnode(0.75,0.56){2.5pt}{A3}      \ncline[linecolor=lightgreen,linewidth=2pt]{A}{A3}
\cnode(1.25,0.56){2.5pt}{A4}      \ncline[linecolor=violet,linewidth=2pt]{A}{A4}
\cnode(0.5,2){2.5pt}{B1}  \ncline[linecolor=green,linewidth=2pt]{B}{B1}
\cnode(1.5,2){2.5pt}{B2}
\psdots[linecolor=green,dotscale=0.6](1.09,2)
\psdots[linecolor=green,dotscale=0.6](1.17,2)
\psdots[linecolor=green,dotscale=0.6](1.25,2)
\psdots[linecolor=green,dotscale=0.6](1.33,2)
\psdots[linecolor=green,dotscale=0.6](1.41,2)

\cnode(0.75,2.43){2.5pt}{B3}      \ncline[linecolor=lightgreen,linewidth=2pt]{B}{B3}
\cnode(1.25,2.43){2.5pt}{B4}      \ncline[linecolor=violet,linewidth=2pt]{B}{B4}

\psarc{->}(1,2){0.25}{270}{180}
\psarc{->}(1,1){0.25}{90}{0}
\psline{->}(1,1)(0.75,1.25)
\psline{->}(1,2)(1.25,1.75)

\rput{90}(0.15,0.89){{\scaleboxto(0,0.18){1st extremity}}}
\rput{90}(0.15,0.89){\psframe[linewidth=0.5pt](-0.4,-0.11)(0.4,0.11)}
\rput{270}(1.85,2){{\scaleboxto(0,0.18){2nd extremity}}}
\rput{270}(1.85,2){\psframe[linewidth=0.5pt](-0.4,-0.11)(0.4,0.11)}

\rput(1,1.5){$e$}
\end{pspicture}}}}

\vskip -3.7mm \noindent extremity, such that each
corresponding edge and face vectors be locked together. Technically,
we represent an edge neighborhood $\eta_e$ by the following structure:

$$
 \eta_e = \bigg\{ 
 \overbrace{
   \overbrace{\bigstrut [\xi_1,\dots,\xi_d]}^{\textrm{Edge vector}},
   \overbrace{\bigstrut [\phi_1,\dots,\phi_d]}^{\textrm{Face
       vector}}
 }^{\textrm{1st extremity}}, 
 \overbrace{
   \overbrace{\bigstrut [\xi'_1,\dots,\xi'_d]}^{\textrm{Edge vector}}, 
   \overbrace{\bigstrut [\phi'_1,\dots \phi'_d]}^{\textrm{Face
       vector}} 
 }^{\textrm{2nd extremity}} 
 \bigg\} 
$$
$$
\textrm{for}~\{\xi_i,\xi'_i\} \subset \mathsf{E}, \quad \textrm{and}~
\{\phi_i,\phi'_i\} \subset \mathsf{F}
$$

\noindent where $\xi_1=\xi'_1=e$, $\phi_1=\phi'_d$ and
$\phi_d=\phi'_1$. The vectors $[\xi_1,\dots,\xi_d]$ and
$[\phi_1,\dots,\phi_d]$ are respectively the edge and
face vectors of the first extremity, and the vectors $[\xi'_1,\dots,\xi'_d]$
and $[\phi'_1,\dots  \phi'_d]$ correspond to the
second extremity.

The {\it color} of an edge neighborhood corresponds to the class of
edges of $e$. The {\it separator} of $\eta_e$, noted
$\textsf{sep}(\eta_e)$, correspond to the pair of classes of faces
separated by $e$, here $(\phi_1,\phi'_1)$. An edge
neighborhood $\eta_e$ colored by $\mathfrak{e}$ is said to be {\it
  coherent} with a pair of vectors $(\xi,\phi)$ if and only if both
edge vectors and face vectors at each extremity of $e$ are isomorphic
to $(\xi,\phi)$.

\medskip

Consider the set of edge neighborhoods of the same color of $\Gamma$. 
As for edge and face vectors, it is possible to define operations on
this set:
\begin{itemize}
\item \textsc{Inversion and Symmetry:} Inversion is the operation
  exchanging both extremities of the edge neighborhood, and
  corresponds to an exchange of the edge vectors and of the faces.
  Symmetry corresponds to the operation of symmetry (defined on the
  edge and face vectors) applied to each extremity of the edge
  neighborhood, while preserving the central edge. 

\item \textsc{Twist or Rearrangement of an extremity:} Let
  $(\xi,\phi)$ be the edge and face vectors associated to one
  extremity of an edge neighborhood $\eta_e$. Any twist or
  rearrangement of $(\xi,\phi)$ that stabilizes the central edge
  leaves the faces separated by $\eta_e$ unchanged and extends
  naturally on the edge neighborhood.
\end{itemize} 

Two edge neighborhoods $\eta_1$ and $\eta_2$ of the same color
$\mathfrak{e}\in\mathsf{E}$ are said to be {\it isomorphic} if and
only if it is possible to transform $\eta_1$ into $\eta_2$ by a
sequence of inversions, symmetries, twists and rearrangements of any
extremity. As was the case for the edge and face vectors, these
operations describe all the possible edge neighborhoods in the
embedding. We therefore select a single representative for each class of
edges in $\Gamma$.

\begin{lem}[Edge neighborhood] \label{lem:edgeneigh}
The edge neighborhood colored by $\mathfrak{e}\in\mathsf{E}$ of
$\Gamma$ is independent of the choice of the embedding of $\Gamma$ and
of the edge it is referring to, up to isomorphism. 
\end{lem}

\begin{prf}
  The separator of a class of edge is independent of the edge
  (Lemma~\ref{lem:separation}). Since finite faces are mapped onto
  finite faces by automorphisms of $\Gamma$, the $2$-connected
  components attached to the extremity of a class of edge are mapped
  onto $2$-connected components by automorphism. Therefore, the
  automorphisms mapping an edge onto another edge correspond to a
  rearrangement of composition of either natural transformations of
  the plane preserving this edge but exchanging its extremities
  (inversion and symmetry) or automorphisms leaving the edge and its
  extremities stable (rearrangements and twists).
\end{prf}

Let $(\xi,\phi)$ be an edge and face vector. Two edges $e,f\in\xi$
labeled by the color $\mathfrak{e} \in \mathsf{E}$ are said to be
equivalent, namely $e\sim f$, if and only if there exists an
isomorphism of $(\xi,\phi)$ mapping $e$ onto $f$. Consider the set of
edges in $\xi$ colored by $\mathfrak{e}$. Then
$\textsf{eq}_{\xi,\phi}(\mathfrak{e})$ is the number of classes of
equivalence inside this set with regard to $\sim$. The uniqueness of the
edge neighborhood implies that $\textsf{eq}_{\xi,\phi}(\mathfrak{e})$
is at least one and at most two. As a matter of fact, each class of
equivalence must correspond to an extremity of the edge neighborhood
colored by $\mathfrak{e}$, and this neighborhood only has two
extremities. 

\begin{exm}\label{exm:edgeneighborhoods}
Let us define edge neighborhoods coherent with the pair of edge and
face vectors described in example~\ref{exm:edgefacevector}. We
represent the edge neighborhood associated to the edge colored by
$\mathfrak{r}$ by the following picture:
\begin{center}
  \mbox{~}
  \hskip 1cm
  \hbox{\rotateleft{\begin{pspicture}(-1.1,-2)(1.1,2)

      \psset{unit=1.6,linewidth=2pt}
      \SpecialCoor

      \pscustom[fillstyle=solid,fillcolor=lightvio,linecolor=violet,%
      linearc=0.1,linewidth=1pt]{
        \pspolygon(-0.1,-0.2)(-0.3,-0.4)(-0.5,-0.4)(-0.5,0.4)(-0.1,0.4)}
      \rput[c]{270}(-0.3,0){$\textcolor{violet}{\gamma}$}
      \pscustom[fillstyle=solid,fillcolor=lightora,linecolor=orange,%
      linearc=0.1,linewidth=1pt]{
        \pspolygon(0.1,-0.4)(0.5,-0.4)(0.5,0.4)(0.3,0.4)(0.1,0.2)}
      \rput[c]{270}(0.3,0){$\textcolor{orange}{\beta}$}
      \cnode(0,-0.5){2pt}{A}
      \cnode(0,0.5){2pt}{B}           \ncline[linecolor=lightvio]{A}{B}
      
      \cnode([angle=275,offset=0.6]A){1pt}{A4} \ncline[linecolor=lightvio]{-c}{A}{A4}
      \cnode([angle=215,offset=0.6]A){1pt}{A3} \ncline[linecolor=lightvio]{-c}{A}{A3}
      \cnode([angle=155,offset=0.6]A){1pt}{A2} \ncline[linecolor=notwhite]{-c}{A}{A2}
      \cnode([angle=095,offset=0.6]A){1pt}{A1} \ncline[linecolor=black]{-c}{A}{A1}  
      
      \pscustom[fillstyle=solid,fillcolor=lightgreen,linecolor=green,%
      linearc=0.1,linewidth=1pt]{
        \pspolygon([angle=130,offset=0.1]A)([angle=105,offset=0.7]A)%
        ([angle=145,offset=0.7]A)}
      \rput[c]{270}([angle=125,offset=0.5]A){\small $\textcolor{green}{\alpha}$}
      \pscustom[fillstyle=solid,fillcolor=lightora,linecolor=orange,%
      linearc=0.1,linewidth=1pt]{
        \pspolygon([angle=190,offset=0.1]A)([angle=165,offset=0.7]A)%
        ([angle=205,offset=0.7]A)}
      \rput[c]{270}([angle=185,offset=0.5]A){\small $\textcolor{orange}{\beta}$}
      \pscustom[fillstyle=solid,fillcolor=lightvio,linecolor=violet,%
      linearc=0.1,linewidth=1pt]{
        \pspolygon([angle=250,offset=0.1]A)([angle=225,offset=0.7]A)%
        ([angle=265,offset=0.7]A)}
      \rput[c]{270}([angle=245,offset=0.5]A){\small $\textcolor{violet}{\beta}$}

      \cnode([angle=275,offset=0.6]B){1pt}{B1} \ncline[linecolor=notwhite]{-c}{B}{B1}
      \cnode([angle=335,offset=0.6]B){1pt}{B2} \ncline[linecolor=black]{-c}{B}{B2}  
      \cnode([angle=035,offset=0.6]B){1pt}{B3} \ncline[linecolor=lightvio]{-c}{B}{B3}
      \cnode([angle=095,offset=0.6]B){1pt}{B4} \ncline[linecolor=lightvio]{-c}{B}{B4}
      
      \pscustom[fillstyle=solid,fillcolor=lightora,linecolor=orange,%
      linearc=0.1,linewidth=1pt]{
        \pspolygon([angle=070,offset=0.1]B)([angle=045,offset=0.7]B)%
        ([angle=085,offset=0.7]B)}
      \rput[c]{270}([angle=065,offset=0.5]B){\small $\textcolor{orange}{\beta}$}
      \pscustom[fillstyle=solid,fillcolor=lightvio,linecolor=violet,%
      linearc=0.1,linewidth=1pt]{
        \pspolygon([angle=010,offset=0.1]B)([angle=345,offset=0.7]B)%
        ([angle=025,offset=0.7]B)}
      \rput[c]{270}([angle=005,offset=0.5]B){\small $\textcolor{violet}{\gamma}$}
      \pscustom[fillstyle=solid,fillcolor=lightgreen,linecolor=green,%
      linearc=0.1,linewidth=1pt]{
        \pspolygon([angle=310,offset=0.1]B)([angle=285,offset=0.7]B)%
        ([angle=325,offset=0.7]B)}
      \rput[c]{270}([angle=305,offset=0.5]B){\small $\textcolor{green}{\alpha}$}

      \psarc[linewidth=1pt]{->}(0,-0.5){0.2}{90}{0}
      \psarc[linewidth=1pt]{->}(0,0.5){0.2}{270}{180}
      
      \rput{270}([angle=275,offset=0.7]B){\tiny $2$}
      \rput{270}([angle=335,offset=0.7]B){\tiny $3$} 
      \rput{270}([angle=035,offset=0.7]B){\tiny $4$}
      \rput{270}([angle=095,offset=0.7]B){\tiny $5$}

      \rput{270}([angle=275,offset=0.7]A){\tiny $5$}
      \rput{270}([angle=215,offset=0.7]A){\tiny $4$}
      \rput{270}([angle=155,offset=0.7]A){\tiny $3$}
      \rput{270}([angle=095,offset=0.7]A){\tiny $2$}

      \rput{270}(-0.9,0.7){\scaleboxto(0,0.15){1st extremity}}
      \rput{270}(-0.9,-0.7){\scaleboxto(0,0.15){2nd extremity}}
  \end{pspicture}}}
  \hfill
  \raise 9.5mm \hbox{{\parbox{8cm}{%
      \begin{tabular}{c@{~:~}c@{~,~}c}
        1st extremity  & 
        $[\mathfrak{r},\mathfrak{g},\mathfrak{b},\mathfrak{r},\mathfrak{r}]$ 
        &
        $[\beta,\alpha,\gamma,\beta,\gamma]$ \\
        2nd extremity & 
        $[\mathfrak{r},\mathfrak{b},\mathfrak{g},\mathfrak{r},\mathfrak{r}]$ 
        & 
        $[\gamma,\alpha,\beta,\gamma,\beta]$ \\
  \end{tabular}}}}
  \hfill\mbox{~}
\end{center}
The fact that the faces colored by $\beta$ are infinite allows for the
transformation of this labeling scheme by twists of either extremity. 
\end{exm}

\begin{exm}
While it could seem that the face vector is superfluous, {\it i.e.}
that the graph could be described simply by its edge neighborhoods
without any face vector, consider the particular case of graphs that
are both vertex-tran\-sitive and edge-transitive, as in the
Figure~\ref{fig:edgeneigh}. The $[5;5;5;5]$ and $[3;7;3;7]$ graphs --
denoted by their type vector -- both belong to this class. Both
possess the same edge neighborhoods and edge vec\-tors. Yet these two
graphs are obviously not isomorphic and their group of automorphisms
are distinct, because the first is face-transitive, and the sec\-ond
possesses two different classes of faces. 
\begin{figure}[hbt]
\begin{center}
\begin{tabular}{cc}
\includegraphics[width=6cm]{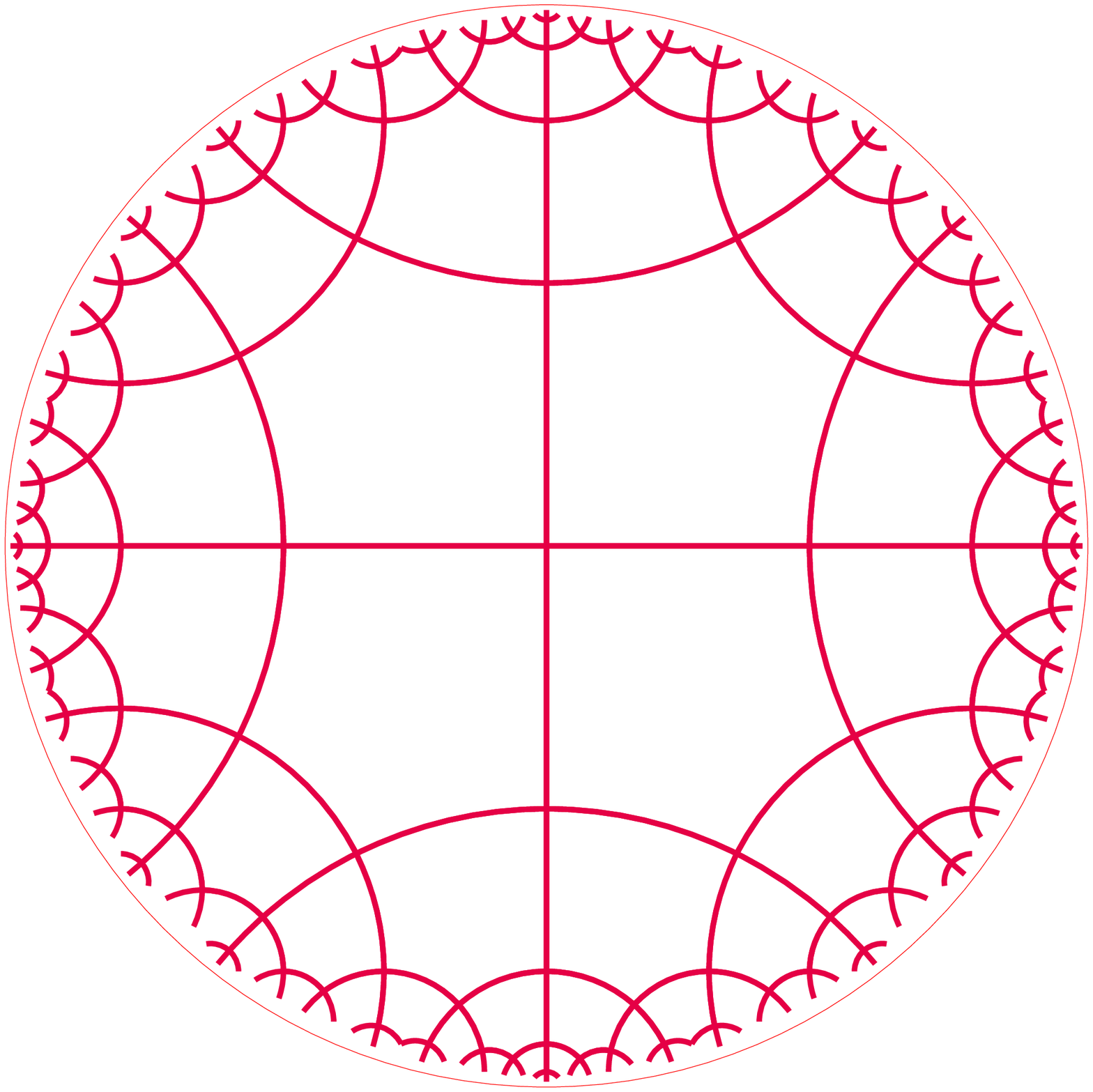}
&
\includegraphics[width=6cm]{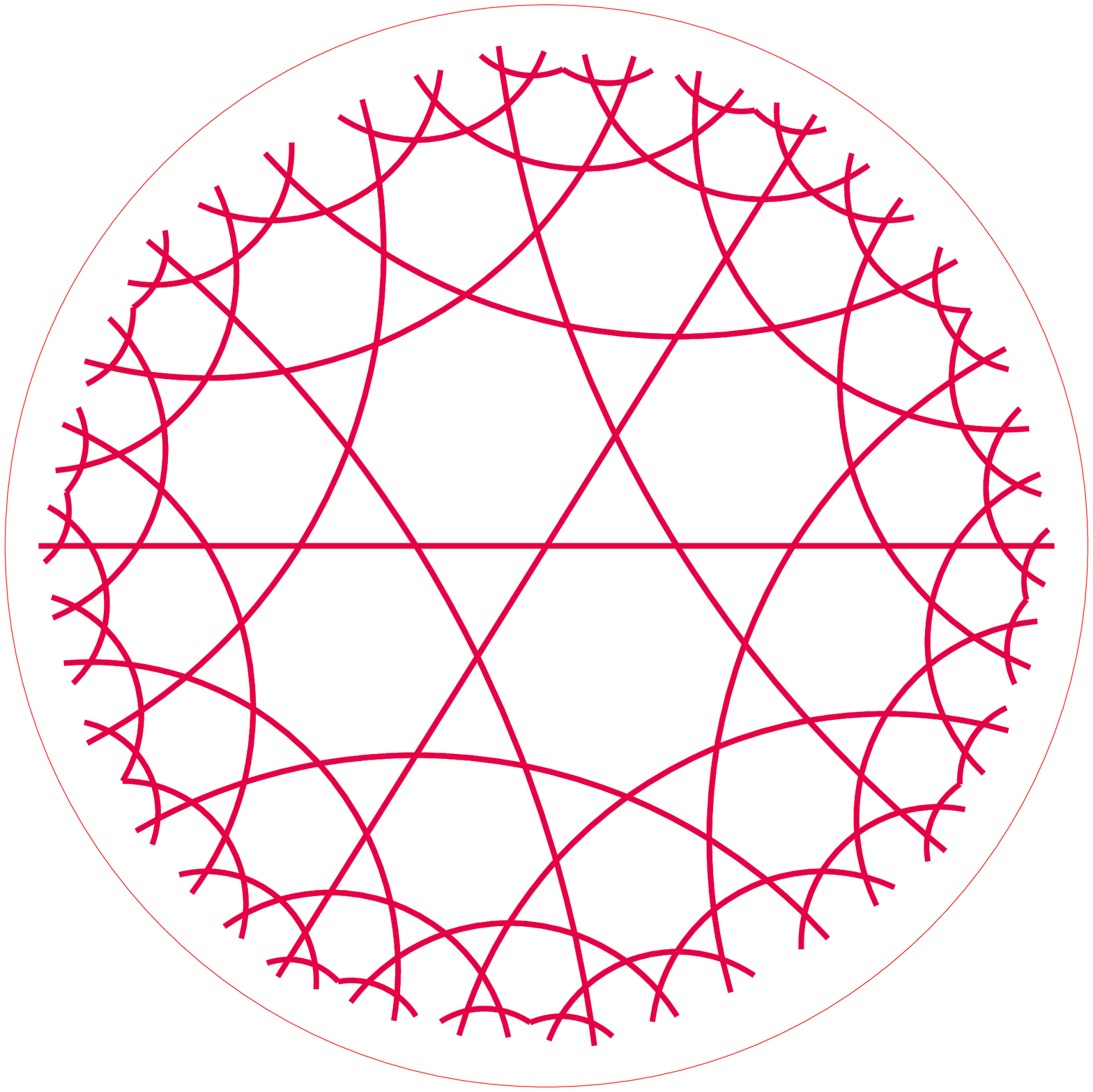}
\end{tabular}
\caption{Non-isomorphic transitive graphs with the same edge
  neighborhoods, assuming that we remove the face vectors from the
  edge neighborhoods.}
\label{fig:edgeneigh}
\end{center}\end{figure}
\end{exm}

Whitney's Theorem \cite{Whitney} states that the finite planar
3-connected graphs have a unique embedding property {\it i.e.} their
dual is uniquely defined. This property was extended to infinite
graphs by Imrich \cite{Imrich}. As a matter of face, when $\Gamma$ is
$3$-connected (while remaining transitive and TLF-planar), all faces
of the graph are finite, and the graph is obviously composed of a
unique $2$-connected component. Therefore the classes of isomorphisms
of the geometrical invariants described in this section contain
neither twists nor rearrangements. This is coherent with the unique
embedding property. Notice that this property holds when $\Gamma$ is
at least $2$-connected, in the case of transitive TLF-planar graphs.
On the other hand, when $\Gamma$ is $1$-separable, these invariants
provide an accurate description of the possible embeddings of the
graph in the plane.

\section{Labeling schemes}
\label{sec:labelingscheme}

Our purpose in this section is to consider the geometrical invariants
of~$\Gamma$ and to prove that they are sufficient to give an exact
description of the graph. The resulting description of the graph is
called a labeling scheme, and extends the notion of labeling scheme
for Cayley graphs, detailed in \cite{Chaboud,DavidCayley}. 

\subsection{Border automaton}

Let $\mathsf{E}$ and $\mathsf{F}$ be two non-intersecting finite
sets of colors. A {\it labeling scheme} of degree $d$ is a 3-tuple
$(\xi,\phi,\eta)$ possessing the following properties:
\renewcommand{\theenumi}{(\roman{enumi})}
\renewcommand{\labelenumi}{(\roman{enumi})}
\begin{enumerate}
\item $\xi \in \mathsf{E}^d$ is an edge vector and $\phi \in
  \mathsf{F}^d$ is a face vector
  ;
\item for each color $\mathfrak{e} \in \mathsf{E}$,
$\textsf{eq}_{\xi,\phi}(\mathfrak{e})$ does not exceed two;
\item for each color $\mathfrak{e}$ in $\xi$, there exists a unique
edge neighborhood $\eta_\mathfrak{e} \in \eta$ of the same color; all
edge neighborhoods in $\eta$ must be coherent with $(\xi, \phi)$ and
if $\textsf{eq}_{\xi,\phi}(\mathfrak{e}) = 2$, then each class of
equivalence must appear on an extremity of $\eta_\mathfrak{e}$.
\end{enumerate}

Given a graph $\Gamma$, then $\xi$ and $\phi$ stand for $\Gamma$'s
edge and face vectors, locked together. The set $\eta$ stands for the
set of edge neighborhoods of $\Gamma$. In the following,
$\eta_{\mathfrak{e}}$ stands for the edge neighborhood in $\eta$
colored by $\mathfrak{e}$. The isomorphisms of labeling schemes are
defined as the isomorphisms of the elements of the scheme.  The
results in the previous sections ensure that for any pair
$(\Gamma,G)$, there exists a labeling scheme $(\xi,\phi,\eta)$
corresponding to the coloring of the vertices and faces associated to
the group $G$ of the graph $\Gamma$, up to isomorphism. Notice that
the condition $(ii)$, along with Lemma~\ref{lem:edgeneigh} constrain
the number of possible labeling schemes.

Consider a labeling scheme $(\xi,\phi,\eta)$. Let $e$ be an element of
$\xi$ and $\eta_e\in\eta$ be the edge neighborhood labeled by the
color of $e$. The operation of {\it gluing $\eta_e$ with $e$} is
possible if and only if there exists an edge neighborhood $\kappa$
isomorphic to $\eta_e$ such that one extremity of $\kappa$ be exactly
equal to the pair $(\xi,\phi)$, with $e\in\xi$ as the central edge of
the neighborhood $\kappa$.

\begin{lem}[Edge reconstruction] \label{lem:reconstruction}
Let $(\xi,\phi,\eta)$ be a labeling scheme, $e \in \xi$ and
$\eta_e \in \eta$ colored by $e$, there exists a unique way to glue
$\eta_e$ with $e$, up to isomorphism of the other extremity of
$\eta_e$.
\end{lem}

\begin{prf}
Suppose that there exists two different ways to glue $\eta_e$ onto
$e$. If each extremity of $\eta_e$ may be glued onto $e$, this means
that all edges colored by $e$ are equivalent with regard to $\sim$. If
only one extremity of $\eta_e$ may be glued onto $e$, then there
exists two classes of equivalence with regard to $\sim$ for the color
of $e$, each of them at one extremity of $\eta_e$. In any case, since
the number of classes of equivalence for the color of $e$ does not
exceed two, that leads to a unique possibility to glue $\eta_e$ onto
$e$, up to isomorphism. 
The existence follows from the fact that every class of equivalence of
every color appears on an extremity of the associated edge
neighborhood. 
\end{prf}

The gluing of edge neighborhoods allows the reconstruction of the
graph. Let us start from an initial graph $\Lambda$ composed of all
edges incident to a central vertex $v$. Suppose that these edges are
labeled accordingly to the edge vector~$\xi$. Obviously, this planar
graph does not include faces for the moment, nevertheless we are
expecting to build a face $\phi_i$ between the edges $\xi_i$ and
$\xi_{i+1}$. By gluing the appropriate edge neighborhood
$\eta_\mathfrak{e}$ onto the edge $\xi_i$, we create $d-1$ new edges
incident to the other extremity of $\xi_i$ labeled such that the
obtained edge neighborhood is isomorphic to $\eta_\mathfrak{e}$. The
other extremities of these new edges correspond to new vertices of the
graph. We will see later how it is possible to close the border of the
faces.

\begin{exm} \label{exm:reconstruction} 
  Consider the labeling scheme defined in Figure~\ref{fig:labscheme1},
  where the set of colors are
  $\mathsf{E}=\{\mathfrak{b},\mathfrak{r},\mathfrak{g}\}$ and
  $\mathsf{F}=\{\alpha,\beta,\gamma\}$. It is based on the
  examples~\ref{exm:edgefacevector} and~\ref{exm:edgeneighborhoods}.
  We associate to each color in $\mathsf{E}$ a unique edge
  neighborhood. The face $\beta$ is supposed to be infinite, thus
  allowing by a twist exchanging the faces colored by $\beta$ that
  $\textsf{eq}_{\xi,\psi}(\mathfrak{r}) = 2$.

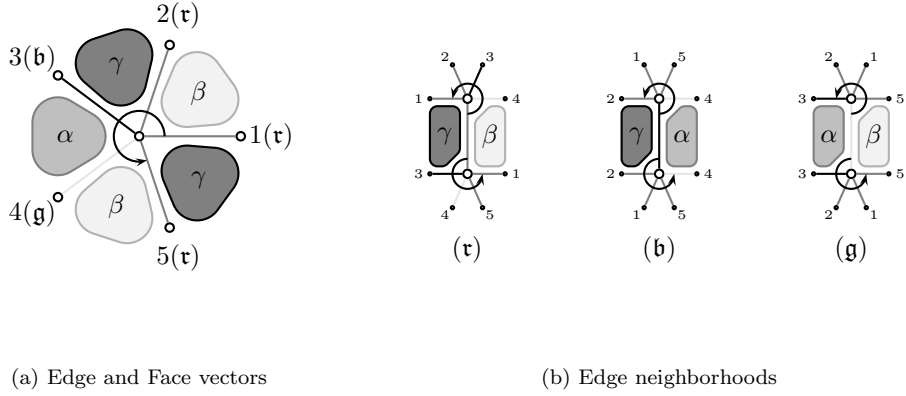
\begin{figure}[htbp]
\begin{center}
\mbox{
  \subfigure[Edge and Face vectors]{\vphantom{\rule[-3mm]{1pt}{4cm}}%
    \begin{pspicture}(-2.5,-2.4)(2.5,2.4)
      \SpecialCoor
      \degrees[5] \psset{unit=1.35}
      
      \pscustom[fillstyle=solid,fillcolor=lightora,linecolor=orange,linearc=0.25]{
        \pspolygon(0.1;0.5)(1.2;0.1)(1.2;0.9)}
      \rput[c](0.725;0.5){$\textcolor{orange}{\beta}$}
      \pscustom[fillstyle=solid,fillcolor=lightvio,linecolor=violet,linearc=0.25]{
        \pspolygon(0.1;1.5)(1.2;1.1)(1.2;1.9)}
      \rput[c](0.725;1.5){$\textcolor{violet}{\gamma}$}
      \pscustom[fillstyle=solid,fillcolor=lightgreen,linecolor=green,linearc=0.25]{
        \pspolygon(0.1;2.5)(1.2;2.1)(1.2;2.9)}
      \rput[c](0.725;2.5){$\textcolor{green}{\alpha}$}
      \pscustom[fillstyle=solid,fillcolor=lightora,linecolor=orange,linearc=0.25]{
        \pspolygon(0.1;3.5)(1.2;3.1)(1.2;3.9)}
      \rput[c](0.725;3.5){$\textcolor{orange}{\beta}$}
      \pscustom[fillstyle=solid,fillcolor=lightvio,linecolor=violet,linearc=0.25]{
        \pspolygon(0.1;4.5)(1.2;4.1)(1.2;4.9)}
      \rput[c](0.725;4.5){$\textcolor{violet}{\gamma}$}

      \cnode(0,0){2pt}{A}
      \cnode(1,0){2pt}{A1}            \ncline[linecolor=lightvio]{A}{A1}
      \rput[c](1.30;0){$1(\textcolor{red}{\mathfrak{r}})$}
      \cnode(0.3,0.9){2pt}{A2}        \ncline[linecolor=lightvio]{A}{A2}
      \rput[c](1.25;1){$2(\textcolor{red}{\mathfrak{r}})$}
      \cnode(-0.8,0.6){2pt}{A3}       \ncline[linecolor=black]{A}{A3}
      \rput[c](1.30;2){$3(\textcolor{black}{\mathfrak{b}})$}
      \cnode(-0.8,-0.6){2pt}{A4}      \ncline[linecolor=notwhite]{A}{A4}
      \rput[c](1.30;3){$4(\textcolor{gold}{\mathfrak{g}})$}
      \cnode(0.3,-0.9){2pt}{A5}       \ncline[linecolor=lightvio]{A}{A5}
      \rput[c](1.25;4){$5(\textcolor{red}{\mathfrak{r}})$}

      \psarc{->}(0,0){0.25}{0}{4}

    \end{pspicture}}
  ~~
  \subfigure[Edge neighborhoods]{\vphantom{\rule[-3mm]{1pt}{4cm}}%
    \raise 2cm \hbox{
      $\begin{array}{ccc} 
        {\begin{pspicture}(-1.1,-1.8)(1.1,1.3)
            
            \pscustom[fillstyle=solid,fillcolor=lightvio,linecolor=violet,linearc=0.1]{
              \pspolygon(-0.1,-0.2)(-0.3,-0.4)(-0.5,-0.4)(-0.5,0.4)(-0.1,0.4)}
            \rput[c](-0.3,0){$\textcolor{violet}{\gamma}$}
            \pscustom[fillstyle=solid,fillcolor=lightora,linecolor=orange,linearc=0.1]{
              \pspolygon(0.1,-0.4)(0.5,-0.4)(0.5,0.4)(0.3,0.4)(0.1,0.2)}
            \rput[c](0.3,0){$\textcolor{orange}{\beta}$}
            
            \cnode(0,-0.5){2pt}{A}
            \cnode(0,0.5){2pt}{B}           \ncline[linecolor=lightvio]{A}{B}
            
            \cnode(0.5,-0.5){1pt}{A4}       \ncline[linecolor=lightvio]{A}{A4}
            \cnode(0.2,-0.95){1pt}{A3}      \ncline[linecolor=lightvio]{A}{A3}
            \cnode(-0.2,-0.95){1pt}{A2}     \ncline[linecolor=notwhite]{A}{A2}
            \cnode(-0.5,-0.5){1pt}{A1}      \ncline[linecolor=black]{A}{A1}  
            
            \cnode(0.5,0.5){1pt}{B1}        \ncline[linecolor=notwhite]{B}{B1}
            \cnode(0.2,0.95){1pt}{B2}       \ncline[linecolor=black]{B}{B2}  
            \cnode(-0.2,0.95){1pt}{B3}      \ncline[linecolor=lightvio]{B}{B3}
            \cnode(-0.5,0.5){1pt}{B4}       \ncline[linecolor=lightvio]{B}{B4}
            
            \psarc{->}(0,-0.5){0.2}{90}{0}
            \psarc{->}(0,0.5){0.2}{270}{180}
            
            \rput(0.65,0.5){\tiny $4$}      \rput(0.65,-0.5){\tiny $1$}
            \rput(0.3,1.05){\tiny $3$}      \rput(0.3,-1.05){\tiny $5$}
            \rput(-0.3,1.05){\tiny $2$}     \rput(-0.3,-1.05){\tiny $4$}
            \rput(-0.65,0.5){\tiny $1$}     \rput(-0.65,-0.5){\tiny $3$}

            \rput(0,-1.5){$(\mathfrak{r})$}
          \end{pspicture}}
        &
        {\begin{pspicture}(-1.1,-1.8)(1.1,1.3)
            
            \pscustom[fillstyle=solid,fillcolor=lightvio,linecolor=violet,linearc=0.1]{
              \pspolygon(-0.1,-0.2)(-0.3,-0.4)(-0.5,-0.4)(-0.5,0.4)(-0.1,0.4)}
            \rput[c](-0.3,0){$\textcolor{violet}{\gamma}$}
            \pscustom[fillstyle=solid,fillcolor=lightgreen,linecolor=green,linearc=0.1]{
              \pspolygon(0.1,-0.4)(0.5,-0.4)(0.5,0.4)(0.3,0.4)(0.1,0.2)}
            \rput[c](0.3,0){$\textcolor{green}{\alpha}$}
            
            \cnode(0,-0.5){2pt}{A}
            \cnode(0,0.5){2pt}{B}           \ncline[linecolor=black]{A}{B}
            
            \cnode(0.5,-0.5){1pt}{A4}       \ncline[linecolor=notwhite]{A}{A4}
            \cnode(0.2,-0.95){1pt}{A3}      \ncline[linecolor=lightvio]{A}{A3}
            \cnode(-0.2,-0.95){1pt}{A2}     \ncline[linecolor=lightvio]{A}{A2}
            \cnode(-0.5,-0.5){1pt}{A1}      \ncline[linecolor=lightvio]{A}{A1}  
            
            \cnode(0.5,0.5){1pt}{B1}        \ncline[linecolor=notwhite]{B}{B1}
            \cnode(0.2,0.95){1pt}{B2}       \ncline[linecolor=lightvio]{B}{B2}   
            \cnode(-0.2,0.95){1pt}{B3}      \ncline[linecolor=lightvio]{B}{B3}
            \cnode(-0.5,0.5){1pt}{B4}       \ncline[linecolor=lightvio]{B}{B4}
            
            \psarc{->}(0,-0.5){0.2}{90}{0}
            \psarc{->}(0,0.5){0.2}{270}{180}
            
            \rput(0.65,0.5){\tiny $4$}      \rput(0.65,-0.5){\tiny $4$}
            \rput(0.3,1.05){\tiny $5$}      \rput(0.3,-1.05){\tiny $5$}
            \rput(-0.3,1.05){\tiny $1$}     \rput(-0.3,-1.05){\tiny $1$}
            \rput(-0.65,0.5){\tiny $2$}     \rput(-0.65,-0.5){\tiny $2$}
            
            \rput(0,-1.5){$(\mathfrak{b})$}
          \end{pspicture}} 
        &
        {\begin{pspicture}(-1.1,-1.8)(1.1,1.3)
            
            \pscustom[fillstyle=solid,fillcolor=lightgreen,linecolor=green,linearc=0.1]{
              \pspolygon(-0.1,-0.2)(-0.3,-0.4)(-0.5,-0.4)(-0.5,0.4)(-0.1,0.4)}
            \rput[c](-0.3,0){$\textcolor{green}{\alpha}$}
            \pscustom[fillstyle=solid,fillcolor=lightora,linecolor=orange,linearc=0.1]{
              \pspolygon(0.1,-0.4)(0.5,-0.4)(0.5,0.4)(0.3,0.4)(0.1,0.2)}
            \rput[c](0.3,0){$\textcolor{orange}{\beta}$}
            
            \cnode(0,-0.5){2pt}{A}
            \cnode(0,0.5){2pt}{B}           \ncline[linecolor=notwhite]{A}{B}
            
            \cnode(0.5,-0.5){1pt}{A4}       \ncline[linecolor=lightvio]{A}{A4}
            \cnode(0.2,-0.95){1pt}{A3}      \ncline[linecolor=lightvio]{A}{A3}
            \cnode(-0.2,-0.95){1pt}{A2}     \ncline[linecolor=lightvio]{A}{A2}
            \cnode(-0.5,-0.5){1pt}{A1}      \ncline[linecolor=black]{A}{A1}   
            
            \cnode(0.5,0.5){1pt}{B1}        \ncline[linecolor=lightvio]{B}{B1}
            \cnode(0.2,0.95){1pt}{B2}       \ncline[linecolor=lightvio]{B}{B2}   
            \cnode(-0.2,0.95){1pt}{B3}      \ncline[linecolor=lightvio]{B}{B3}
            \cnode(-0.5,0.5){1pt}{B4}       \ncline[linecolor=black]{B}{B4}
            
            \psarc{->}(0,-0.5){0.2}{90}{0}
            \psarc{->}(0,0.5){0.2}{270}{180}
            
            \rput(0.65,0.5){\tiny $5$}      \rput(0.65,-0.5){\tiny $5$}
            \rput(0.3,1.05){\tiny $1$}      \rput(0.3,-1.05){\tiny $1$}
            \rput(-0.3,1.05){\tiny $2$}     \rput(-0.3,-1.05){\tiny $2$}
            \rput(-0.65,0.5){\tiny $3$}     \rput(-0.65,-0.5){\tiny $3$}
            
            \rput(0,-1.5){$(\mathfrak{g})$}
          \end{pspicture}}
      \end{array}$} 
  }
}
\end{center}
\caption{A labeling scheme of degree $5$ with its components. }
\label{fig:labscheme1}
\end{figure}


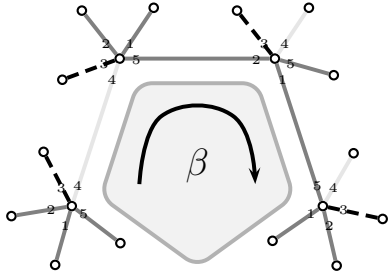
\begin{figure}[ht]
  \begin{center} 
    {\begin{pspicture}(-1.6,-1.5)(1.6,2.4)
        \psset{unit=2.7,linewidth=1.5pt}
        \SpecialCoor
        \degrees[5]

        \pscustom[fillstyle=solid,fillcolor=lightora,linecolor=orange,linearc=0.1]{%
          \pspolygon(0.5;3.75)(0.5;4.75)(0.5;0.75)(0.5;1.75)(0.5;2.75)
        }
        \rput[c](0.,0.){\Large $\textcolor{orange}{\beta}$}

        \pscurve{<-}(0.3;4.75)(0.3;0.75)(0.3;1.75)(0.3;2.75)

        \cnode[linewidth=0.8pt](0.65;4.75){2pt}{A0}      
        \cnode[linewidth=0.8pt](0.65;0.75){2pt}{A1}     
        \ncline[linecolor=lightvio]{A0}{A1}
        \cnode[linewidth=0.8pt](0.65;1.75){2pt}{A2}
        \ncline[linecolor=lightvio]{A1}{A2} 
        \cnode[linewidth=0.8pt](0.65;2.75){2pt}{A3}
        \ncline[linecolor=notwhite]{A2}{A3} 

        \cnode[linewidth=0.8pt]([angle=3.6,offset=0.3]A1){2pt}{B1}       
        \ncline[linecolor=lightvio]{A1}{B1}
        \cnode[linewidth=0.8pt]([angle=4.6,offset=0.3]A1){2pt}{B2}       
        \ncline[linecolor=notwhite]{A1}{B2}
        \cnode[linewidth=0.8pt]([angle=0.6,offset=0.3]A1){2pt}{B3}       
        \ncline[linecolor=black,linestyle=dashed]{A1}{B3}

        \cnode[linewidth=0.8pt]([angle=4.6,offset=0.3]A2){2pt}{B1}       
        \ncline[linecolor=lightvio]{A2}{B1}
        \cnode[linewidth=0.8pt]([angle=0.6,offset=0.3]A2){2pt}{B2}       
        \ncline[linecolor=lightvio]{A2}{B2}
        \cnode[linewidth=0.8pt]([angle=1.6,offset=0.3]A2){2pt}{B3}       
        \ncline[linecolor=black,linestyle=dashed]{A2}{B3}

        \cnode[linewidth=0.8pt]([angle=0.45,offset=0.3]A3){2pt}{B1}      
        \ncline[linecolor=black,linestyle=dashed]{A3}{B1}
        \cnode[linewidth=0.8pt]([angle=1.45,offset=0.3]A3){2pt}{B2}      
        \ncline[linecolor=lightvio]{A3}{B2}
        \cnode[linewidth=0.8pt]([angle=2.35,offset=0.3]A3){2pt}{B3}      
        \ncline[linecolor=lightvio]{A3}{B3}
        \cnode[linewidth=0.8pt]([angle=3.3,offset=0.3]A3){2pt}{B4}       
        \ncline[linecolor=lightvio]{A3}{B4}

        \cnode[linewidth=0.8pt]([angle=1.8,offset=0.3]A0){2pt}{B1}
        \ncline[linecolor=lightvio]{A0}{B1} 
        \cnode[linewidth=0.8pt]([angle=2.75,offset=0.3]A0){2pt}{B2}
        \ncline[linecolor=lightvio]{A0}{B2} 
        \cnode[linewidth=0.8pt]([angle=3.65,offset=0.3]A0){2pt}{B3}
        \ncline[linecolor=black,linestyle=dashed]{A0}{B3} 
        \cnode[linewidth=0.8pt]([angle=4.65,offset=0.3]A0){2pt}{B4}
        \ncline[linecolor=notwhite]{A0}{B4} 

        \rput(-0.65,-0.3){\tiny$1$}\rput(-0.72,-0.22){\tiny$2$}
        \rput(-0.67,-0.11){\tiny$3$}\rput(-0.59,-0.10){\tiny$4$}
        \rput(-0.56,-0.24){\tiny$5$}

        \rput(-0.42,0.42){\tiny$4$}\rput(-0.46,0.50){\tiny$3$}
        \rput(-0.45,0.6){\tiny$2$}\rput(-0.33,0.6){\tiny$1$}
        \rput(-0.3,0.52){\tiny$5$}

        \rput(0.42,0.42){\tiny$1$}\rput(0.46,0.50){\tiny$5$}
        \rput(0.43,0.6){\tiny$4$}\rput(0.33,0.6){\tiny$3$}
        \rput(0.29,0.52){\tiny$2$}

        \rput(0.65,-0.3){\tiny$2$}\rput(0.72,-0.22){\tiny$3$}
        \rput(0.67,-0.11){\tiny$4$}\rput(0.59,-0.10){\tiny$5$}
        \rput(0.56,-0.24){\tiny$1$}

      \end{pspicture}}
  \end{center} 
\caption{Successive gluings of the edge neighborhoods along the border
  of the face colored by $\beta$.}
\label{fig:successive}
\end{figure}

Let us try to glue the edge neighborhoods consecutively, while
following the edges constituting the border of the orange face (cf.
Figure~\ref{fig:successive}). We begin by gluing the golden edge
neighborhood $\eta_\mathfrak{g}$ onto the unique golden $\mathfrak{g}$
edge belonging to the edge vector. Having glued two red edge
neighborhoods $\eta_\mathfrak{r}$, we can ourselves continue the
process indefinitely, by gluing red edge neighborhoods along the
border of the face. If this labeling scheme corresponds to an existing
graph $\Gamma$, then this process describes the border of a face
colored by $\beta$ in~$\Gamma$.
\end{exm}

\parbox{\textwidth}{\begin{lem}[Description of the faces] Consider a
    graph $\Gamma$ and its labeling scheme $(\xi,\phi,\eta)$. It is
    possible to build a finite state automaton and an operation onto
    that automaton allowing to construct the borders of the faces of~$\Gamma$.
\end{lem}}

\begin{prf}
  Consider an automaton built over the alphabet $A$. A language
  $L \subset A^\star$ acts naturally over the set of states of the
  automaton. Therefore, we consider the following partition of the
  states of the automaton: two states $u$ and $v$ are equivalent if
  and only if there exists $l\in L$ such that it is possible to start
  from the state $u$, and reach the state $v$ by reading the word $l$
  on the automaton. Now we will build an automaton by describing its
  set of states and the language acting on these states. 

  A {\it configuration} of a labeling scheme $(\xi,\phi,\eta)$ is a
  pair $(\mathcal{C},b)$ where $\mathcal{C}$ is a class of
  equivalence for $\sim$ in $(\xi,\phi)$ and $b\in\{+,-\}$ is a
  direction of rotation. A configuration represents a block of edges
  (separated by infinite faces) attached to a vertex, an edge inside
  this component and a direction expressing in which way the block is
  embedded in the plane. Two configurations are said to be
  \textit{equivalent} if $(i)$ they correspond to two blocks, $(ii)$
  it is possible to map the first block onto the second by an
  isomorphism of edge and face vectors and $(iii)$ this map sends the
  edge of the first configuration onto the edge of the second
  configuration, and maps the direction of rotation accordingly. 
  
  For a given labeling scheme, the number of configurations is finite,
  bounded by $2d$. These configurations define the states of our
  automaton. Let us define the following relations between
  configurations:
\begin{itemize}
\item The \textsf{next} relation describes whether a configuration
comes next to another one in the edge vector, given a direction of
rotation: $(\mathcal{C},b) \overset{\textsf{\tiny next}}{\longrightarrow}
(\mathcal{C}',b)$ if and only if there exist $(\bar{\xi},\bar{\phi})$
isomorphic to $(\xi,\phi)$ such that if $\bar{\xi}_i$ (the $i^\textrm{th}$
element of the vector $\xi$) belongs to $\mathcal{C}$, the
edge next to $\bar{\xi}_i$ according to the direction $b$ (if $b$ is
positive, that is $\bar{\xi}_{i+1}$, otherwise that is $\bar{\xi}_{i-1}$)
belongs to $\mathcal{C}'$.

\item The \textsf{inv} relation describes how the configuration is
modified when we cross the corresponding edge: $(\mathcal{C},b)
\overset{\textsf{\tiny inv}}{\longrightarrow} (\mathcal{C}',b')$ if
and only both configurations correspond to the same color of edge
$\mathfrak{e}$ , and an edge neighborhood isomorphic to
$\eta_{\mathfrak{e}}\in\eta$ has $(\mathcal{C},b)$ and
$(\mathcal{C}',b')$ at its extremities.
\end{itemize}

Given these two relations, it is possible to build a finite state
automaton on the set of configurations. This automaton is called the
{\it border automaton} associated to this labeling scheme. By
Lemma~\ref{lem:reconstruction}, it is connected. The \textsf{inv}
relation is involutive, while the \textsf{next} relation may have more
than one successor, therefore this automaton is non-deterministic in
general. Determining the border of a face $\mathcal{F}$ is only a
matter of reading the infinite word $(\textsf{next}\cdot
\textsf{inv})^\omega$, with starting state a configuration such that
the face $\mathcal{F}$ appears between this configuration and the
configuration \textsf{next} to it in the automaton. Such a word is an
orbit of the automaton under the action of $(\textsf{next}\cdot
\textsf{inv})$. 
\end{prf}

For a given labeling scheme, we can associate to an element of the
face vector -- or a face -- an orbit of the automaton. The colors of
the faces are supposed to distinguish the classes of the faces under
the group of automorphism. Two finite faces are said to be {\it
  equivalent} if they have the same orbit or orbits that correspond to
the same border read in opposite directions. If two faces of the same
color are not equivalent, then the labeling scheme is said to be {\it
  invalid}. Labeling schemes resulting from TLF-planar graphs are
always valid. This property ensures that given two faces of the same
color, there exists an automorphism mapping the first onto the
second. 

The orbits of the automaton can be classified into two categories :
the cyclic orbits and the acyclic ones. Cyclic orbits correspond to
faces that can be ``closed'', meaning that their border is periodic.
Consider the orbit containing the $i$-th edge of $\xi$, with positive
direction of rotation. If this orbit is cyclic, then we define $k_i$
as the size of this orbit, otherwise $k_i=\infty$. Let $(a_i),
i\in[1;d]$ be a set of formal letters such that $a_i=a_j$ if and only
if the i-th and j-th face of $\phi$ are of the same color. The vector
whose elements are the $k_i a_i$ (or simple $\infty$ if $k_i$ is
infinite) is called the {\it primitive type vector} of
$(\xi,\phi,\eta)$. A type vector $[l_1,\dots,l_d]$ is said to be {\it
  valid} with regard to that labeling scheme if and only if there
exists a valuation of the $(a_i)$ in $\Nat$ such that $\forall i, l_i
= k_i a_i$ and all values in the vector are greater than three.

\subsection{Examples of construction}

\begin{exm} \label{exm:threeconnex}
  Consider the labeling scheme defined in Figure~\ref{fig:labscheme2},
  with degree~$5$, $\textsf{E} =
  \{\mathfrak{b},\mathfrak{r},\mathfrak{g}\}$ and $\textsf{F} = \{
  \alpha,\beta, \gamma\}$. With each color in $\textsf{E}$ we
  associate a unique edge neighborhood. Let us try to compute the
  associated border automaton.  Even if $|\textsf{E}|=3$, there exist
  $10$ different configurations, one for each edge and direction of
  rotation. Since the $\textsf{next}$ operation corresponds to a
  rotation of the edge vector, we represent the border automaton with
  two cycles of length $5$. On the Figure~\ref{fig:bordaut2}, \textsf{next}
  edges appear in black, while \textsf{inv} edges appear in dashed
  gray lines:

\begin{figure}
  \begin{center}
    {\begin{pspicture}(-2.2,-2)(2.4,2)
        \SpecialCoor
        \degrees[5] \psset{unit=1.3}

        \pscustom[fillstyle=solid,fillcolor=lightgreen,linecolor=green,linearc=0.25]{
          \pspolygon(0.1;0.5)(1.2;0.1)(1.2;0.9)}
        \rput[c](0.725;0.5){$\textcolor{green}{\beta}$}
        \pscustom[fillstyle=solid,fillcolor=lightvio,linecolor=violet,linearc=0.25]{
          \pspolygon(0.1;1.5)(1.2;1.1)(1.2;1.9)}
        \rput[c](0.725;1.5){$\textcolor{blue}{\alpha}$}
        \pscustom[fillstyle=solid,fillcolor=lightgreen,linecolor=green,linearc=0.25]{
          \pspolygon(0.1;2.5)(1.2;2.1)(1.2;2.9)}
        \rput[c](0.725;2.5){$\textcolor{green}{\beta}$}
        \pscustom[fillstyle=solid,fillcolor=lightgreen,linecolor=green,linearc=0.25]{
          \pspolygon(0.1;3.5)(1.2;3.1)(1.2;3.9)}
        \rput[c](0.725;3.5){$\textcolor{green}{\beta}$}
        \pscustom[fillstyle=solid,fillcolor=lightora,linecolor=orange,linearc=0.25]{
          \pspolygon(0.1;4.5)(1.2;4.1)(1.2;4.9)}
        \rput[c](0.725;4.5){$\textcolor{red}{\gamma}$}

        \cnode(0,0){2pt}{A}
        \cnode(1,0){2pt}{A1}            \ncline[linecolor=lightvio]{A}{A1}
        \rput[c](1.3;0){$1(\mathfrak{r})$}
        \cnode(0.3,0.9){2pt}{A2}        \ncline[linecolor=black]{A}{A2}
        \rput[c](1.2;1){$2(\mathfrak{b})$}
        \cnode(-0.8,0.6){2pt}{A3}       \ncline[linecolor=black]{A}{A3}
        \rput[c](1.3;2){$3(\mathfrak{b})$}
        \cnode(-0.8,-0.6){2pt}{A4}      \ncline[linecolor=notwhite]{A}{A4}
        \rput[c](1.3;3){$4(\mathfrak{g})$}
        \cnode(0.3,-0.9){2pt}{A5}       \ncline[linecolor=lightvio]{A}{A5}
        \rput[c](1.2;4){$5(\mathfrak{r})$}

        \psarc{->}(0,0){0.25}{0}{4}

      \end{pspicture}}
    ~~
    \raise 2cm \hbox{
      $\begin{array}{ccc} 
        {\begin{pspicture}(-1.1,-1.8)(1.2,1.3)

            \pscustom[fillstyle=solid,fillcolor=lightgreen,linecolor=green,linearc=0.1]{
              \pspolygon(-0.1,-0.2)(-0.3,-0.4)(-0.5,-0.4)(-0.5,0.4)(-0.1,0.4)}
            \rput[c](-0.3,0){$\textcolor{green}{\beta}$}
            \pscustom[fillstyle=solid,fillcolor=lightora,linecolor=orange,linearc=0.1]{
              \pspolygon(0.1,-0.4)(0.5,-0.4)(0.5,0.4)(0.3,0.4)(0.1,0.2)}
            \rput[c](0.3,0){$\textcolor{red}{\gamma}$}

            \cnode(0,-0.5){2pt}{A}
            \cnode(0,0.5){2pt}{B}           \ncline[linecolor=lightvio]{A}{B}

            \cnode(0.5,-0.5){1pt}{A4}       \ncline[linecolor=lightvio]{A}{A4}
            \cnode(0.2,-0.95){1pt}{A3}      \ncline[linecolor=notwhite]{A}{A3}
            \cnode(-0.2,-0.95){1pt}{A2}     \ncline[linecolor=black]{A}{A2}
            \cnode(-0.5,-0.5){1pt}{A1}      \ncline[linecolor=black]{A}{A1}  

            \cnode(0.5,0.5){1pt}{B1}        \ncline[linecolor=lightvio]{B}{B1}
            \cnode(0.2,0.95){1pt}{B2}       \ncline[linecolor=black]{B}{B2}  
            \cnode(-0.2,0.95){1pt}{B3}      \ncline[linecolor=black]{B}{B3}
            \cnode(-0.5,0.5){1pt}{B4}       \ncline[linecolor=notwhite]{B}{B4}

            \psarc{->}(0,-0.5){0.2}{90}{0}
            \psarc{->}(0,0.5){0.2}{270}{180}

            \rput(0.65,0.5){\tiny $1$}      \rput(0.65,-0.5){\tiny $5$}
            \rput(0.3,1.05){\tiny $2$}      \rput(0.3,-1.05){\tiny $4$}
            \rput(-0.3,1.05){\tiny $3$}     \rput(-0.3,-1.05){\tiny $3$}
            \rput(-0.65,0.5){\tiny $4$}     \rput(-0.65,-0.5){\tiny $2$}

            \rput(0,-1.5){$(\mathfrak{r})$}

          \end{pspicture}}
        &
        {\begin{pspicture}(-1.1,-1.8)(1.2,1.3)

            \pscustom[fillstyle=solid,fillcolor=lightvio,linecolor=violet,linearc=0.1]{
              \pspolygon(-0.1,-0.2)(-0.3,-0.4)(-0.5,-0.4)(-0.5,0.4)(-0.1,0.4)}
            \rput[c](-0.3,0){$\textcolor{blue}{\alpha}$}
            \pscustom[fillstyle=solid,fillcolor=lightgreen,linecolor=green,linearc=0.1]{
              \pspolygon(0.1,-0.4)(0.5,-0.4)(0.5,0.4)(0.3,0.4)(0.1,0.2)}
            \rput[c](0.3,0){$\textcolor{green}{\beta}$}

            \cnode(0,-0.5){2pt}{A}
            \cnode(0,0.5){2pt}{B}           \ncline[linecolor=black]{A}{B}

            \cnode(0.5,-0.5){1pt}{A4}       \ncline[linecolor=lightvio]{A}{A4}
            \cnode(0.2,-0.95){1pt}{A3}      \ncline[linecolor=lightvio]{A}{A3}
            \cnode(-0.2,-0.95){1pt}{A2}     \ncline[linecolor=notwhite]{A}{A2}
            \cnode(-0.5,-0.5){1pt}{A1}      \ncline[linecolor=black]{A}{A1}  

            \cnode(0.5,0.5){1pt}{B1}        \ncline[linecolor=notwhite]{B}{B1}
            \cnode(0.2,0.95){1pt}{B2}       \ncline[linecolor=lightvio]{B}{B2}   
            \cnode(-0.2,0.95){1pt}{B3}      \ncline[linecolor=lightvio]{B}{B3}
            \cnode(-0.5,0.5){1pt}{B4}       \ncline[linecolor=black]{B}{B4}

            \psarc{->}(0,-0.5){0.2}{90}{0}
            \psarc{->}(0,0.5){0.2}{270}{180}

            \rput(0.65,0.5){\tiny $4$}      \rput(0.65,-0.5){\tiny $1$}
            \rput(0.3,1.05){\tiny $5$}      \rput(0.3,-1.05){\tiny $5$}
            \rput(-0.3,1.05){\tiny $1$}     \rput(-0.3,-1.05){\tiny $4$}
            \rput(-0.65,0.5){\tiny $2$}     \rput(-0.65,-0.5){\tiny $3$}
            
            \rput(0,-1.5){$(\mathfrak{b})$}
          \end{pspicture}}
        &
        {\begin{pspicture}(-1.1,-1.8)(1.2,1.3)

            \pscustom[fillstyle=solid,fillcolor=lightgreen,linecolor=green,linearc=0.1]{
              \pspolygon(-0.1,-0.2)(-0.3,-0.4)(-0.5,-0.4)(-0.5,0.4)(-0.1,0.4)}
            \rput[c](-0.3,0){$\textcolor{green}{\beta}$}
            \pscustom[fillstyle=solid,fillcolor=lightgreen,linecolor=green,linearc=0.1]{
              \pspolygon(0.1,-0.4)(0.5,-0.4)(0.5,0.4)(0.3,0.4)(0.1,0.2)}
            \rput[c](0.3,0){$\textcolor{green}{\beta}$}

            \cnode(0,-0.5){2pt}{A}
            \cnode(0,0.5){2pt}{B}           \ncline[linecolor=notwhite]{A}{B}

            \cnode(0.5,-0.5){1pt}{A4}       \ncline[linecolor=black]{A}{A4}
            \cnode(0.2,-0.95){1pt}{A3}      \ncline[linecolor=black]{A}{A3}
            \cnode(-0.2,-0.95){1pt}{A2}     \ncline[linecolor=lightvio]{A}{A2}
            \cnode(-0.5,-0.5){1pt}{A1}      \ncline[linecolor=lightvio]{A}{A1}   

            \cnode(0.5,0.5){1pt}{B1}        \ncline[linecolor=lightvio]{B}{B1}
            \cnode(0.2,0.95){1pt}{B2}       \ncline[linecolor=lightvio]{B}{B2}   
            \cnode(-0.2,0.95){1pt}{B3}      \ncline[linecolor=black]{B}{B3}
            \cnode(-0.5,0.5){1pt}{B4}       \ncline[linecolor=black]{B}{B4}

            \psarc{->}(0,-0.5){0.2}{90}{0}
            \psarc{->}(0,0.5){0.2}{270}{180}

            \rput(0.65,0.5){\tiny $5$}      \rput(0.65,-0.5){\tiny $3$}
            \rput(0.3,1.05){\tiny $1$}      \rput(0.3,-1.05){\tiny $2$}
            \rput(-0.3,1.05){\tiny $2$}     \rput(-0.3,-1.05){\tiny $1$}
            \rput(-0.65,0.5){\tiny $3$}     \rput(-0.65,-0.5){\tiny $5$}

            \rput(0,-1.5){$(\mathfrak{g})$}

          \end{pspicture}}
      \end{array}$} 
  \end{center}
  \caption{Another labeling scheme of degree 5}
  \label{fig:labscheme2}
\end{figure}
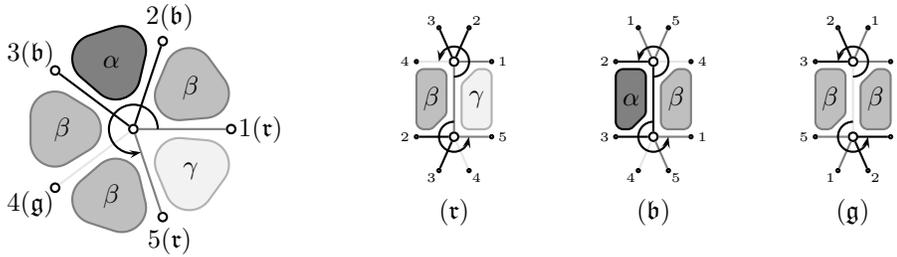

\begin{figure}[htbp]
\begin{center}
\newcounter{iia}
\begin{pspicture}(-2,-1.7)(2,1.8)

\psset{unit=0.8}
\SpecialCoor
\degrees[5]
\multido{\ia=2+1}{5}{%
\setcounter{iia}{\ia}
\addtocounter{iia}{-1}
\psline[linestyle=none](0;0)(1.5;\ia)
\pscircle(2;\ia){2pt}           
\psline[arrowsize=0.2]{<-}(2;\ia)(2;\theiia)
\rput[c](2.4;\ia){$\theiia^+$}
\pscircle(1.4;\ia){2pt}         
\psline[arrowsize=0.2]{->}(1.4;\ia)(1.4;\theiia)
\rput[c](1;\ia){$\theiia^-$}}

\psset{linestyle=dashed}
\pscurve[linecolor=lightgray](2;2)(2.4;1.5)(2;1)
\pscurve[linecolor=lightgray](1.4;2)(1.8;1.5)(1.4;1)
\pscurve[linecolor=lightgray](2;0)(2.3;0.1)(2.3;-0.1)(2;0)
\pscurve[linecolor=lightgray](1.4;0)(1.7;0.1)(1.7;-0.1)(1.4;0)
\pscurve[linecolor=lightgray](2;3)(2.4;3.5)(2;4)
\pscurve[linecolor=lightgray](1.4;3)(1.8;3.5)(1.4;4)

\pscircle*[linecolor=lightgreen](2;2){0.1}
\pscircle*[linecolor=lightgreen](1.4;1){0.1}
\pscircle*[linecolor=violet](2;3){0.1}
\pscircle*[linecolor=violet](2;0){0.1}
\pscircle*[linecolor=violet](2;1){0.1}
\pscircle*[linecolor=violet](1.4;0){0.1}
\pscircle*[linecolor=violet](1.4;2){0.1}
\pscircle*[linecolor=violet](1.4;4){0.1}
\pscircle[linecolor=turquoise,fillstyle=solid,fillcolor=white](2;4){0.1}
\pscircle[linecolor=turquoise,fillstyle=solid,fillcolor=white](1.4;3){0.1}
\end{pspicture}
~~
\begin{pspicture}(-1.5,-1.7)(2,1.8)

\psframe[linecolor=lightgray](-0.7,-1.5)(3.3,1.5)
\psset{arrowinset=0.1}

\pnode(-0.3,1){A} \nccircle[angleA=300,angleB=60,nodesepA=0.1]{->}{A}{0.2}
\pscircle*[linecolor=lightgreen](-0.3,1){0.1}
\rput(-0.11,1.105){\tiny $2$}
\rput[l](0.4,1){First orbit}

\pnode(-0.3,0){A} \nccircle[angleA=300,angleB=60,nodesepA=0.1]{->}{A}{0.2}
\pscircle*[linecolor=violet](-0.3,0){0.1}
\rput(-0.11,0.105){\tiny $3$}
\rput[l](0.4,0){Second orbit}

\pnode(-0.3,-1){A} \nccircle[angleA=300,angleB=60,nodesepA=0.1]{->}{A}{0.2}
\pscircle[linecolor=turquoise,fillstyle=solid,fillcolor=white](-0.3,-1){0.1}
\rput(-0.11,-.895){\tiny $2$}
\rput[l](0.4,-1){Third orbit}
\end{pspicture}
\end{center}
\caption{The border automaton associated to the labeling scheme
  in Figure~\ref{fig:labscheme2}.}
\label{fig:bordaut2}
\end{figure}
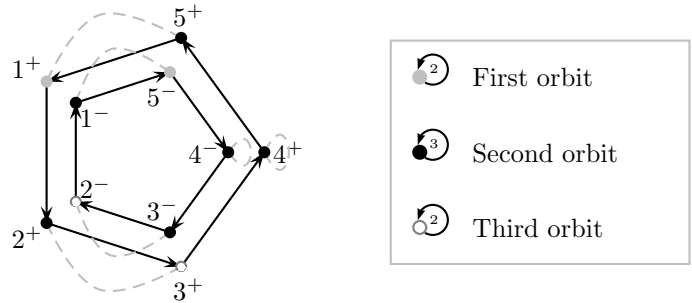

The possible orbits are therefore $[1]$, $[2,4,5]$, and $[3]$. The
first one corresponds to faces with red borders $\mathfrak{r}^\star$,
and the third one to faces with blue borders $\mathfrak{b}^\star$. The
others have border $(\mathfrak{bgr})^\star$. We have three possible
classes of borders of the faces, corresponding to each orbit of the
automaton and to each color in $\mathsf{F}$.  The corresponding
primitive type vector is: $[3n,m,3n,3n,p]$ for $(n,m,p)\in \Nat$. On
Figure~\ref{fig:example1}, there is an planar embedding in the
hyperbolic plane of a graph possessing this labeling scheme and type
vector $[3,4,3,3,5]$. The faces {\Large $\alpha$} correspond
squares and the faces {\Large $\gamma$} to pentagons, while
the faces with 3-colored borders are triangles. Notice that this graph
possesses a trivial stabilizer and is therefore a Cayley graph.

  \begin{figure}[htbp]
    \begin{center}
      
      \mbox{~}\hfill
      \hbox{\begin{minipage}{7.4cm}
        \caption{Example of a 3-connected TLF-planar vertex-transitive graph,
          with type vector $[5;3;4;3;3]$, associated to the labeling
          sche\-me described in the Example~4.}
        \label{fig:example1}
      \end{minipage}}
      \hfill
      \begin{minipage}{6cm}
        \begin{center}
          {\begin{pspicture}(-2.5,-2.8)(2.5,2.8)
              \rput(0.3,0){\includegraphics[width=7cm]{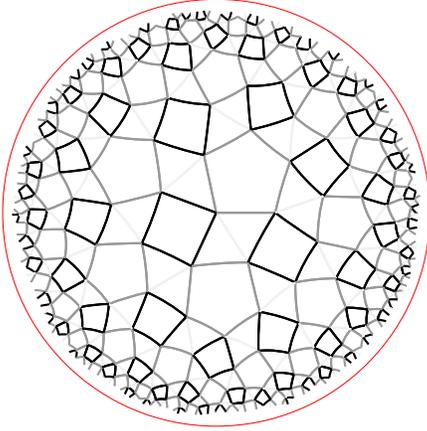}}
            \end{pspicture}}
          \end{center}
      \end{minipage}
      \hfill \mbox{~}

    \end{center}
  \end{figure}

\end{exm}

\begin{figure}[htbp]
    \begin{center}
      
      \mbox{~}\hfill
      \begin{minipage}{7.3cm}
        \caption{Example of a 1-separable TLF-planar vertex-transitive graph,
          with type vector $[3;\infty;3;4;\infty]$, associated to the
          labeling scheme described in the Example~5.}
        \label{fig:example2}
      \end{minipage}
      \hfill
      \begin{minipage}{6cm}
        \begin{center}
          {\begin{pspicture}(-2.5,-2.8)(2.5,2.8)
              \rput(0.3,0){\includegraphics[width=7cm]{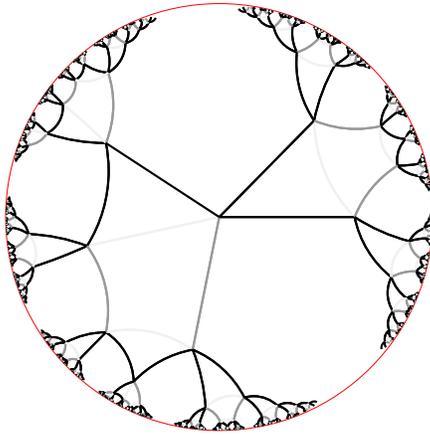}}
            \end{pspicture}}
          \end{center}
      \end{minipage}
      \hfill \mbox{~}

    \end{center}
  \end{figure}

\begin{exm} \label{exm:aperiodic}
Consider now the case of example~\ref{exm:reconstruction}
page~\pageref{exm:reconstruction}, where
$\mathsf{E}=\{\mathfrak{b},\mathfrak{r},\mathfrak{g}\}$ and
$\mathsf{F}=\{\alpha,\beta,\gamma\}$. As in the previous example, we
associate to each color in $\mathsf{E}$ a unique edge
neighborhood. The labeling scheme is valid with regard to our
definitions. When we assume that the face $\beta$ is infinite, we come
down with the border automaton appearing on Figure~\ref{fig:bordaut1}.

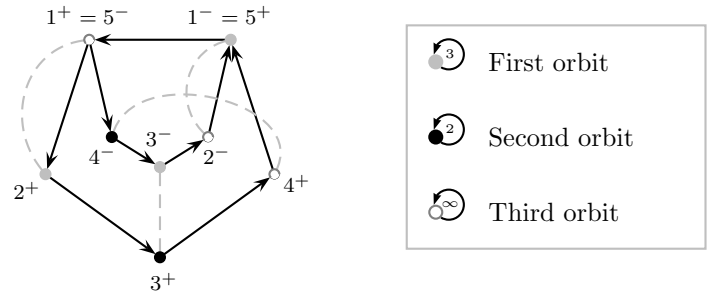
\begin{figure}[htbp]
  \begin{center}
    \begin{pspicture}(-2,-1.9)(2,1.7)

\psset{unit=0.8}
\SpecialCoor
\degrees[5]

\cnode(2;0.75){2pt}{A1}           
\cnode(2;1.75){2pt}{A2}           
\cnode(2;2.75){2pt}{A3}           
\cnode(2;3.75){2pt}{A4}           
\cnode(2;4.75){2pt}{A5}           

\ncline[arrowsize=0.2]{->}{A1}{A2}
\ncline[arrowsize=0.2]{->}{A2}{A3}
\ncline[arrowsize=0.2]{->}{A3}{A4}
\ncline[arrowsize=0.2]{->}{A4}{A5}
\ncline[arrowsize=0.2]{->}{A5}{A1}

\rput([angle=0.2,offset=0.4]A1){\footnotesize $1^-=5^+$}
\rput([angle=0.2,offset=0.4]A2){\footnotesize $1^+=5^-$}
\rput([angle=2,offset=0.4]A3){\footnotesize $2^+$}
\rput([angle=2.8,offset=0.4]A4){\footnotesize $3^+$}
\rput([angle=3.6,offset=0.4]A5){\footnotesize $4^+$}

\cnode(-0.8,0){2pt}{B3}           
\cnode(0,-0.5){2pt}{B4}           
\cnode(0.8,0){2pt}{B5}  

\ncline[arrowsize=0.2]{->}{A2}{B3}
\ncline[arrowsize=0.2]{->}{B3}{B4}
\ncline[arrowsize=0.2]{->}{B4}{B5}
\ncline[arrowsize=0.2]{->}{B5}{A1}

\rput(-0.95,-0.3){\footnotesize $4^-$}
\rput(0,0){\footnotesize $3^-$}
\rput(0.95,-0.3){\footnotesize $2^-$}

\psset{linestyle=dashed}
\ncline[linecolor=lightgray]{B4}{A4}
\nccurve[linecolor=lightgray,angleA=1,angleB=0.9,ncurv=0.9]{B3}{A5}
\nccurve[linecolor=lightgray,angleA=1.9,angleB=2.7,ncurv=0.9]{B5}{A1}
\nccurve[linecolor=lightgray,angleA=1.8,angleB=2.6,ncurv=0.9]{A3}{A2}

\pscircle*[linecolor=lightgreen](2;2.75){0.1}
\pscircle*[linecolor=lightgreen](0,-0.5){0.1}
\pscircle*[linecolor=lightgreen](2;0.75){0.1}
\pscircle*[linecolor=violet](2;3.75){0.1}
\pscircle*[linecolor=violet](-0.8,0){0.1}
\pscircle[linecolor=turquoise,fillstyle=solid,fillcolor=white](2;1.75){0.1}
\pscircle[linecolor=turquoise,fillstyle=solid,fillcolor=white](2;4.75){0.1}
\pscircle[linecolor=turquoise,fillstyle=solid,fillcolor=white](0.8,0){0.1}

\end{pspicture}
~~
\begin{pspicture}(-1.5,-1.9)(2,1.7)

\psframe[linecolor=lightgray](-0.7,-1.5)(3.3,1.5)
\psset{arrowinset=0.1}

\pnode(-0.3,1){A} \nccircle[angleA=300,angleB=60,nodesepA=0.1]{->}{A}{0.2}
\pscircle*[linecolor=lightgreen](-0.3,1){0.1}
\rput(-0.11,1.105){\tiny $3$}
\rput[l](0.4,1){First orbit}

\pnode(-0.3,0){A} \nccircle[angleA=300,angleB=60,nodesepA=0.1]{->}{A}{0.2}
\pscircle*[linecolor=violet](-0.3,0){0.1}
\rput(-0.11,0.105){\tiny $2$}
\rput[l](0.4,0){Second orbit}

\pnode(-0.3,-1){A} \nccircle[angleA=300,angleB=60,nodesepA=0.1]{->}{A}{0.2}
\pscircle[linecolor=turquoise,fillstyle=solid,fillcolor=white](-0.3,-1){0.1}
\rput(-0.11,-0.895){\tiny $\infty$}
\rput[l](0.4,-1){Third orbit}
\end{pspicture}
\end{center}
\caption{The border automaton associated to the labeling scheme in
  Figure~\ref{fig:labscheme1}.}
\label{fig:bordaut1}
\end{figure}

The first orbit corresponds to the face $\gamma$,
which is bordered by $(\mathfrak{rrb})^\star$. The second orbit
corresponds to the face $\alpha$, bordered by
$(\mathfrak{bg})^\star$. More interesting is the case of the 
face $\beta$. Notice that due to the possible twists,
the \textsf{next} operator is not deterministic.

This leads to faces bordered by aperiodic words from
$\mathfrak{r}^\omega \mathfrak{gr}^\omega$. There does not exist any
embedding of the graph can avoid these aperiodic faces (cf.
Fig.~\ref{fig:example2}), and all infinite faces either have a unique
yellow edge in their border or none at all. The associated primitive
type vector is given by $[\infty,3n,2m,\infty,3n]$ for values of
$(m,n) \in \Nat$, which leads to 1-separable graphs such as the one on
the right. For Cayley graphs, it is always possible to find an
embedding where the border of the faces are periodic
\cite{DavidCayley}. Hence, this set of graphs contains no Cayley
graphs, because of this aperiodic border property.
\end{exm}

\subsection{Realization}

Every TLF-planar vertex-transitive graph possesses a valid labeling
scheme and a vector type that is valid regarding to that scheme.
Consider now the converse of this result: given a general valid
labeling scheme, and a valid type vector, is it possible to produce a
vertex-transitive graph that has the same labeling scheme and type
vector ?

Our goal in this section consists in building specific embeddings of
vertex-transitive planar graphs with particular geometrical
properties. Depending on the graph, we select an appropriate geometry:
Euclidean, spherical or hyperbolic. As long as our graphs are TLF-planar,
embeddings in spherical geometry will lead to finite graphs. Infinite
faces thus can only occur in Euclidean and hyperbolic geometry. The
following theorem constructs embeddings that are tilings of the plane
by regular polygons.

\begin{thm}[Existence] \label{thm:existence}
Given a labeling scheme $(\xi,\phi,\eta)$, and a valid type vector
$[k_1,\dots,k_d]$, there exists a TLF-planar
vertex-transi\-tive graph possessing this scheme and type
vector. Moreover, all faces of the embedding are regular polygons.
\end{thm}

\begin{prf}
Consider any point in the plane. We are going to evaluate the interior
angle of a regular polygon of side length $l$, with $k_i$ sides, and
then the total angle corresponding to all the polygons in the type
vector must be equal to $2\pi$. The following values result from
simple trigonometry in the different geometries:  
\renewcommand{\arraystretch}{1.5}
$$ 
\theta_i(l) = \left\{
\begin{array}{cl}
\textrm{Spherical :}&
\strut 2\arcsin \left( \frac{\cos(\pi/k_i)}{\cos(l/2)}\right) \\
\textrm{Euclidean :}&
\strut \frac{(k_i-2)}{k_i}\pi\\
\textrm{Hyperbolic :}&
\strut 2\arcsin \left( \frac{\cos(\pi/k_i)}{\cosh(l/2)}\right)
\end{array} 
\right. \quad \textrm{and} \quad \sum_{i=1}^d \theta_i(l) =
2\pi\quad(1)$$
The equation $(1)$ determines the choice of the
geometry : let $\Sigma$ be the sum of the angles in the Euclidean
plane. If $\Sigma=2\pi$, we choose the Euclidean geometry, and any
value is possible for $l$ (the corresponding graphs will be
homothetic). If $\Sigma<2\pi$, we choose the spherical geometry,
whereas if $\Sigma>2\pi$ we select the hyperbolic geometry. In any
case, there exists an unique solution of the equation $(1)$ that
determines a unique value for the length $l$, and consequently for the
angles $\theta_i$\footnote{Except for the following type vectors :
  $[3,3,p\geq 5]$, $[3,4,p\geq 6]$ and $[3,5,p\geq 9]$. There does not
  exist a labeling scheme validating any of these type vectors (cf
  p~\pageref{page:tv}).}. These values allow us to draw all the
regular polygons that correspond to our type vector around a given
point of the plane.

Let $\epsilon$ be a point in the plane. By induction, we build a
planar locally finite graph $\Gamma_n$, with a central vertex
$\epsilon$, such that all vertices at distance $\leq n$ from
$\epsilon$ have degree $d$. The graph $\Gamma_1$ is composed of the
glued polygons of the first case, centered onto $\epsilon$, but
restricted to the edges incident to $\epsilon$. The labeling of
$\Gamma_1$ is chosen isomorphic to $(\xi,\phi)$, such that the face
$\phi_i$ corresponds to a regular polygon with $k_i$ sides. 
Suppose now that we have built $\Gamma_n$. Consider the finite set of
vertices of $\Gamma_n$ at distance $n$ of degree less than $d$. We
shall build the remaining edges with geodesics, such that the length
of the geodesic be $l$, and the angle at the vertex corresponding to
the face $\phi_i$ be equal to~$\theta_i(l)$.

The labeling scheme describes how the edges must be labeled : take a
vertex $s$ and an edge $e$ colored by $\mathfrak{e}$ incident to $s$.
According to Lemma~\ref{lem:reconstruction}, there exists a unique way
to glue the edge neighborhood $\eta_\mathfrak{e}$ onto $e$, up to
isomorphism. The choice of the edge neighborhood is not crucial,
because the isomorphisms leave the angles of the faces constant. The
construction of $\Gamma_{n+1}$ is just a matter of gluing the edge
neighborhoods on the edges possessing an extremity at distance $n+1$
from $\epsilon$, by choosing an adequate edge neighborhood isomorphic
to $\eta_\mathfrak{e}$.

This construction can not lead to intersecting edges. In fact; all
edges have the same length, and inside a given face, all angles are
the same. Therefore the construction of a face ultimately leads to a
regular polygon in the plane. This answers the problem of the closure
of the faces. 

\begin{center}
{\begin{pspicture}(-3,-2)(3,2)

\pscustom[fillstyle=solid,fillcolor=notwhite,linestyle=none]{%
  \pspolygon(0,0.5)(0.707,1.207)(0,1.914)(-0.707,1.207)}
\pscustom[fillstyle=solid,fillcolor=notwhite,linestyle=none]{%
  \pspolygon(0,-0.5)(0.707,-1.207)(0,-1.914)(-0.707,-1.207)}
\pscustom[fillstyle=solid,fillcolor=notwhite,linestyle=none]{%
  \pspolygon(-2.414,0.5)(-1.707,1.207)(-0.707,1.207)(0,0.5)%
(0,-0.5)(-0.707,-1.207)(-1.707,-1.207)(-2.414,-0.5)}

\psarc[linecolor=red](0,0.5){0.2}{270}{45}
\uput{0.25}[337.5](0,0.5){\footnotesize $\theta_l$}
\psarc[linecolor=red](0,-0.5){0.2}{315}{90}
\uput{0.25}[22.5](0,-0.5){\footnotesize $\theta_l$}
\psarc[linecolor=red](0.707,1.207){0.2}{225}{0}
\uput{0.25}[292.5](0.707,1.207){\footnotesize $\theta_l$}
\psarc[linecolor=red](0.707,-1.207){0.2}{0}{135}
\uput{0.25}[67.5](0.707,-1.207){\footnotesize $\theta_l$}
\psarc[linecolor=red](1.707,-1.207){0.2}{45}{180}
\uput{0.25}[112.5](1.707,-1.207){\footnotesize $\theta_l$}
\psarc[linecolor=red](1.707,1.207){0.2}{180}{315}
\uput{0.25}[247.5](1.707,1.207){\footnotesize $\theta_l$}

\cnode(0,0.5){2pt}{A} 
\cnode(0,-0.5){2pt}{AA} \ncline{A}{AA} 
\pcline[linecolor=red,offset=7pt]{<->}(0,-0.5)(0,0.5)
\lput*{:U}{\footnotesize $l$}

\cnode(0.707,1.207){2pt}{A1} \ncline{A}{A1}
\pcline[linecolor=red,offset=7pt]{<->}(0,0.5)(0.707,1.207)
\lput*{:U}{\footnotesize $l$}

\cnode(1.707,1.207){2pt}{A2} \ncline{A1}{A2}
\pcline[linecolor=red,offset=7pt]{<->}(0.707,1.207)(1.707,1.207)
\lput*{:U}{\footnotesize $l$}

\cnode(2.414,0.5){2pt}{C2} \ncline{A2}{C2}
\pcline[linecolor=red,offset=7pt]{<->}(1.707,1.207)(2.414,0.5)
\lput*{:U}{\footnotesize $l$}

\cnode(0.707,-1.207){2pt}{A3} \ncline{AA}{A3}
\pcline[linecolor=red,offset=7pt]{<->}(0.707,-1.207)(0,-0.5)
\lput*{:U}{\footnotesize $l$}

\cnode(1.707,-1.207){2pt}{A4} \ncline{A3}{A4}
\pcline[linecolor=red,offset=7pt]{<->}(1.707,-1.207)(0.707,-1.207)
\lput*{:U}{\footnotesize $l$}

\cnode(2.414,-0.5){2pt}{A5} \ncline{A4}{A5}
\pcline[linecolor=red,offset=7pt]{<->}(2.414,-0.5)(1.707,-1.207)
\lput*{:U}{\footnotesize $l$}

\cnode(3.121,-1.207){2pt}{A6} \ncline{A5}{A6}
\cnode(2.414,-1.914){2pt}{A7} \ncline{A4}{A7}

\cnode(-0.707,1.207){2pt}{AA1} \ncline{A}{AA1}
\cnode(0,1.914){2pt}{B1} \ncline{AA1}{B1}\ncline{B1}{A1}
\cnode(-1.707,1.207){2pt}{AA2} \ncline{AA1}{AA2}
\cnode(-0.707,-1.207){2pt}{AA3} \ncline{AA}{AA3}
\cnode(0,-1.914){2pt}{B2} \ncline{AA3}{B2} \ncline{B2}{A3}
\cnode(-1.707,-1.207){2pt}{AA4} \ncline{AA3}{AA4}
\cnode(-2.414,-0.5){2pt}{AA5} \ncline{AA4}{AA5} 

\cnode(-3.121,-1.207){2pt}{B1} \ncline{AA5}{B1}
\cnode(-2.414,0.5){2pt}{AA6} \ncline{AA6}{AA5} \ncline{AA6}{AA2}

\cnode[linecolor=gray,linestyle=dashed](2.414,0.2){2pt}{C} 
\ncline[linecolor=gray,linestyle=dashed]{C}{A5}
\rput(2.6,0.25){\textcolor{red}{?}}

\rput(1.207,0){{$\mathcal{F}$}}
  \end{pspicture}}
~
{\begin{pspicture}(-0.5,-2)(5,2)

\rput[l](1.4,-0.425){Length of the edges}
\cnode(0,-0.5){2pt}{Y1}
\cnode(1.1,-0.5){2pt}{Y3}    \ncline{Y1}{Y3}
\pcline[linecolor=red,offset=7pt]{<->}(0,-0.5)(1.1,-0.5)
\lput*{:U}{\footnotesize $l$}

\rput[l](1.4,0.425){Angle inside face $\mathcal{F}$}
\psclip{\psframe[linestyle=none](0,0.15)(1.1,0.85)}
\cnode(0.55,0.7){2pt}{Y2}
\psarc[linecolor=red](0.55,0.7){0.2}{225}{315}
\uput{0.25}[270](0.55,0.7){\footnotesize $\theta_l$}
\pnode(0,0){Y5}   \ncline{Y2}{Y5}
\pnode(1.1,0){Y6} \ncline{Y2}{Y6}

\endpsclip
\end{pspicture}}
\end{center}

Hence the limit graph $\Gamma$ is well defined. This graph is
vertex-transitive: the construction is independent of the starting
vertex, and applying this construction with two different starting
vertices creates an automorphism mapping the first vertex on the
second. The labeling of the edges entails that the automorphism group
is transitive. The automorphisms stabilizing a vertex exactly
correspond to the classes of equivalence of the configurations.
\end{prf}

\noindent
\parbox{\textwidth}{\begin{thm}[Coherence]
If $\Gamma$ is a TLF-planar vertex-transi\-tive graph that
possesses a labeling scheme $(\xi,\phi,\eta)$ and a type vector
$[k_1,\dots,k_d]$, then the preceding construction yields a graph
isomorphic to~$\Gamma$.
\end{thm}}

Therefore, for any vertex-transi\-tive graph obeying our conditions,
it is possible to embed the graph in a particular geometry of the
plane with regular polygons. Moreover, automorphisms of the graph map
finite faces onto isometric faces. Consequently, in the case of the
2-connected graphs, every automorphism of $\Gamma$ extends into an
isometry of the plane. This result appears in Babai \cite{BabaiGrowth}
in the case of 3-connected graphs, and is here extended to the larger
case of the TLF-planar graphs. Notice that the case of $1$-separable
graphs is more complex: even is the faces of the graph are regular
polygons, example~\ref{exm:aperiodic} shows that there exists graphs
for which it is impossible to find an embedding where automorphisms
may be realized by isometries of the plane.

A direct consequence of this result pertains to the growth-rate of the
TLF-planar graphs. When $\Gamma$ is $2$-separable, its structure is
arborescent and its growth-rate is either linear or exponential. On
the other hand, when $\Gamma$ is at least $3$-connected, the embedding
of $\Gamma$ given by the Theorem~\ref{thm:existence} is
quasi-isometric to the plane inside which it is defined. Therefore the
growth-rate of the graph is either quadratic or exponential. 

\subsection{Combinatorial approach}

Consider the problem of finding all the valid labeling schemes for
graphs of degree $d$. The number of colors in $\textsf{E}$ and
$\textsf{F}$ is bounded by $d$. Therefore the number of possible edge
and face vectors is bounded by $d^{d-1}$, up to rotation, and the
number of edge neighborhoods for a given color by $2d$. It is
therefore possible to enumerate all possible labeling schemes, and
then to extract the valid schemes by computing the border automaton.
The particular properties of labeling schemes allow to
drastically reduce this theoretical upper bound. 

\begin{thm}[Enumeration] \label{thm:enum}
Given a number $d\geq 2$, it is possible to enumerate all TLF-planar
transitive graphs having internal degree $d$, along with their
labeling scheme and primitive type vector.
\end{thm}

A possible question lays in determining all possible type vectors for
TLF-planar vertex-transitive graphs. We enumerate the graphs of a given
degree, and eliminate the redundant ones by testing if they belong to
the same class of isomorphism.  For example, the possible type vectors
for vertex-transitive graphs of degree $3$ are:
\label{page:tv}
$$ \bigg\{ [n,n,n], [n,2m,2m], [2n,2m,2p] \bigg\}$$ 

\dots for values of $n,m,p$ such that the faces of the graphs are at
least triangles. The set of planar graphs obtained is presented in
appendix, and extends the vertex-transitive graphs presented in
\cite{Tilings}.  Notice also that this set is strictly larger than the
set of possible Cayley graphs of the same degree, containing in
particular the graph associated to the dodecahedron, with type vector
$[5;5;5]$ and the icosidodecahedron, with type vector $[3;5;3;5]$,
both of which are not Cayley graphs. More generally, this is the case
of all tilings with type vector $\{[6n\pm 1,6n\pm 1,6n\pm 1]\}$. This
is an example o an infinite family of TLF-planar transitive graphs
which are not Cayley.

More precisely, for a given labeling scheme $(\xi,\phi,\eta)$ and
valid type vector, we have the possibility to check whether the
associated unlabeled graph is Cayley or not, by enumerating all Cayley
graphs having the same type vector \cite{DavidCayley}, computing the
labeling schemes of these transitive graphs and comparing them to
$(\xi,\phi,\eta)$:

\begin{cor}[Cayley checking]
If $\Gamma$ is a TLF-planar transitive graph, then it is
decidable whether $\Gamma$ is the Cayley graph of a group or not, and
obtain an enumeration and a description of the groups having $\Gamma$
as a Cayley graph. 
\end{cor}

Another side-result is the possibility to enumerate for a given graph
$\Gamma$ all the groups of automorphisms acting transitively on the
set of vertices of $\Gamma$. This raises the question on finding
graphs which are vertex-transitive but not Cayley, {\it i.e.} who do
not possess a group of automorphism acting simply transitively on its
vertices. Consider the stabilizer of a vertex of $\Gamma$: it must be
a finite group of isometries of the plane fixing the vertex, hence it
is either cyclic or dihedral. If this stabilizer does not contain any
rotation, then the subgroup of direct isometries of $\Aut$
acts simply transitively on the vertices of $\Gamma$, therefore such
graphs must possess rotations in their stabilizers. 

The $\textsf{multiply}(k)$ operation on a labeling scheme
$(\phi,\xi,\eta)$ of degree $d$ consists in building another labeling
scheme $(\phi',\xi',\eta')$ of degree $k\times d$ defined by:
\begin{itemize}
\item $\phi'$ and $\xi'$ consist of $k$ copies of $\phi$ and $\xi$,
{\it i.e.} $\phi'_i = \phi_{(i \mod d)}$, and $\xi'_i = \xi_{(i \mod d)}$;
\item each edge neighborhood $\eta'_{\mathfrak{e}}$ is a copy of
$\eta_{\mathfrak{e}}$ where to the edge and face vectors at each
extremity have been replaced by $(\xi',\phi')$. 
\end{itemize}

The set of configurations of the automaton remains unchanged by this
operation, as is the construction of the border automaton. This
operation allows to build labeling schemes with stabilizers containing
rotations. Moreover, this operation is invertible: for a labeling
scheme whose stabilizer possesses rotations, it is possible to extract
a labeling scheme without rotations by an operation of division. 

\begin{lem}[Taxonomy of prime degree graphs]
  Let $\Gamma$ be a 3-connected planar ver\-tex-transi\-tive graph of
  prime degree $d$.  Either $\Gamma$ is a face and edge-transitive
  graph with type vector $[k;\dots;k]$, or it is a Cayley graph.
\end{lem}

\begin{prf}
  Consider the group of automorphisms of $\Gamma$ and the stabilizer
  $G_s$ of a vertex $s$ of $\Gamma$. Since the graph can be embedded
  in the plane by regular polygons, so that the automorphisms map the
  finite faces onto finite faces, $G$ is a group of isometries of the
  plane and $G_s$ is a finite group of isometries of the plane leaving
  $s$ stable. If $G_s$ contains a non-trivial rotation, since the
  degree of $\Gamma$ is prime, the graph is edge and face-transitive.
  On the contrary, if $G_s$ contains no rotation, then it has a single
  non-trivial element which is a symmetry. Then the
  subgroup of the direct isometries in $G$ acts simply and
  transitively on the vertices of $\Gamma$. Therefore $\Gamma$ is a
  Cayley graph. 
\end{prf}

This analysis points out the classes of graphs that are likely not to
be Cayley graphs. For the lesser degrees, these graphs are either
face-transitive $[k;\dots;k]$, or edge-transitive
$[k_1,k_2,\dots,k_1,k_2]$. There exists an example of graph of degree
$9$ belonging to this class that is neither face-transitive nor
edge-transitive.

\section*{Discussion}

The geometrical invariants of TLF-planar graphs allowed here the
description of the graphs by their local properties, in the case of
vertex-transitive graphs. Nevertheless, there exists different
directions where we could extend this study. First, the dual graph of
a transitive planar graph, is not transitive, but cofinite. That is to
say that there exists only a finite number of orbits of the vertices
under the action of the group of automorphisms. Even if these graphs
in general do not respect the same geometrical properties as the
transitive graphs, it would be interesting to study this family and
their embeddings. We could then question the possibility of extending
the representation by labeling schemes to cofinite graphs. Second, the
key property of local finiteness is rather restrictive: our graphs
possess one accumulation point at most.  Yet Levinson showed that the
number of accumulation points, if superior to one, is either two or
infinite. The analysis of the embeddings of planar graphs with two or
more accumulation points could lead to a geometrical representation
for a larger class of vertex-transitive planar graphs.

\bibliographystyle{alpha}
\bibliography{prethese}

\newpage

\appendix
\newpage 
\mbox{}
\thispagestyle{empty}
\vfill
\begin{center}
{\Huge \bf Appendix}
\end{center}
\vfill
\newpage
\addtocounter{page}{-1}
\section{Enumerations of vertex-transitive graphs}

In the following sections, we enumerate all locally finite planar
vertex-transitive graphs of small inner degree (from 3 to
4). Transitive graphs of degree $2$ correspond ot cyclic graphs. For
the larger degrees, we enumerate all possible labeling schemes, each
one corresponding to an enumerable family of planar vertex-transitive
graphs. Some of these families may have the same representative in
terms of unlabeled graphs, but the groups of automorphisms are
distinct. These families are classified depending on whether their
borders are periodic (P) or aperiodic (A). The following table
displays the number of such classes :

\renewcommand{\arraystretch}{1.2}
$$
\begin{array}{|ccc||ccc||} \hline
\textrm{Degree} & \textrm{~~P~~}  & \textrm{~~A~~}  & \textrm{Degree}  &
\textrm{~~P~~}  & \textrm{~~A~~}  \\ \hline 
1      &   0    &   0      & 4      & 52     & 1       \\ \hline
2      &   1    &   0      & 5      & 174    & >1       \\ \hline
3      &   16   &   0      & 6      & 775    & >1       \\ \hline
\end{array}
$$ 

\medskip
For each labeling scheme, we give the primitive type vector, and a
description of the graph as a list containing the classes of edges and
the border of the faces. Each class of edge is numbered $a_i^k$, where
$k$ is a boolean indicating whether the associated edge neighborhood
reverses the direction of rotation of the edge vector or not. Each
border of the faces is a word on the edge classes corresponding to an
orbit in the border automaton repeated a given number of times. 

\end{document}